\documentclass[prd, twocolumn,noshowpacs,preprintnumbers,amsmath,amssymb,
nofootinbib]{revtex4}

\usepackage{amsmath, amssymb}
\usepackage{graphicx, epic, eepic}% Include figure files
\usepackage{dcolumn}% Align table columns on decimal point
\usepackage{bm}% bold math

  \renewcommand{\a}{\alpha}
  \renewcommand{\b}{\beta}
  
  \renewcommand{\d}{\delta}

    \newcommand{\g}{\gamma}
    \newcommand{\G}{\Gamma}

    \newcommand{\p}{\varphi}
  
    \newcommand{\s}{\sigma}
  
  \newcommand{\ve}{\varepsilon} 
    \newcommand{\x}{\xi}

  \newcommand{\RR}{\mathbb{R}}

  \newcommand{\half}{\frac{1}{2}}

\def\p1{\phantom{1}}
\def\IR{{\hbox{{\rm I}\kern-.2em\hbox{\rm R}}}}

\begin{document}

\preprint{HUTP-05/A0010}

\title{A Higgs Mechanism for Gravity} 

\author{Ingo Kirsch}
\affiliation{\\ Jefferson Laboratory of Physics, Harvard University\\  
Cambridge, MA 02138, USA }

%\date{}% It is always \today, today,
             %  but any date may be explicitly specified

\begin{abstract} 

In this paper we elaborate on the idea of an emergent spacetime which
arises due to the dynamical breaking of diffeomorphism invariance in
the early universe. In preparation for an explicit symmetry breaking
scenario, we consider nonlinear realizations of the group of
analytical diffeomorphisms which provide a unified description of
spacetime structures. We find that gravitational fields, such as the
affine connection, metric and coordinates, can all be interpreted as
Goldstone fields of the diffeomorphism group. We then construct a
Higgs mechanism for gravity in which an affine spacetime evolves into
a Riemannian one by the condensation of a metric.  The symmetry
breaking potential is identical to that of hybrid inflation but with
the non-inflaton scalar extended to a symmetric second rank tensor.
This tensor is required for the realization of the metric as a Higgs
field.  We finally comment on the role of Goldstone coordinates as a
dynamical fluid of reference.
\end{abstract}


\maketitle

%%%%%%%%%%%%%%%%%%%%%%%%%%%%%%%%%%%%%%%%%%%%%%%%%%%%%%%%%%%%%%%%%%%%%
\section{Introduction}
%%%%%%%%%%%%%%%%%%%%%%%%%%%%%%%%%%%%%%%%%%%%%%%%%%%%%%%%%%%%%%%%%%%%%

Recent discoveries in cosmology have shown that general relativity 
is most likely incomplete. In particular the high degree of
homogeneity and isotropy of the universe can only be understood by
\mbox{supplementing} Einstein's theory with an inflationary scenario. Also
the accelerating expansion of the universe by dark energy might hint
to a further modification of general relativity.

Most of these additions to general relativity introduce new fields in
a quite {\em ad hoc} way. For instance, inflationary models typically
postulate one or two scalar fields which drive a rapid expansion of
the early universe.  Even in general relativity the metric
is not derived from any underlying symmetry principle. An
exception are gauge theories of gravity in which the existence of a
connection is justified by the gauging. Also in ghost
condensation~\cite{Nima} the dynamical field appears as the Goldstone
boson of a spontaneously broken time diffeomorphism symmetry. However,
in many other cases the group theoretical origin of the gravitational
fields remains unclear.

In this paper we show that the existence of most of these fields can
be understood in terms of Goldstone bosons which arise in a rapid
symmetry breaking phase shortly after the Big Bang. 

We assume that at the beginning of the universe all spacetime
structures were absent and consider the universe as a Hilbert space
${\cal H}$ accommodating spinor and tensor representations of the
analytic diffeomorphism group $\overline{\textit{Diff}}\,(n, \RR)$.
Unlike in general relativity spinors and tensors are representations
of the same covering group $\overline{\textit{Diff}}\,(n, \RR)$ whose
existence has been shown in \cite{Neeman}. Since spinor
representations of $\overline{\textit{Diff}}\,(n, \RR)$ are
necessarily infinite-dimensional \cite{Neeman}, matter would look
quite exotic at this stage. Here we are however not so much interested
in the structure of these representations. For our purposes, it is
enough to suppose that they do exist.

We further assume that in a series of spontaneous symmetry breakings,
the transformation group $H$ of states in the Hilbert space
collapsed down to the Lorentz group.  We suggest that the symmetry
breaking sequence is given by the group inclusion
\begin{align}
  \overline{\textit{Diff}}\,(n, \RR) \stackrel{TR}{\supset} 
 \overline{\textit{Diff}}_0(n, \RR)  \stackrel{NL}{\supset}
  \overline{GL}(n, \RR)
  \stackrel{SD}\supset \overline{SO}(1,n-1)  \nonumber
\end{align}
where $\overline{\textit{Diff}}_0(n, \RR)$ is the homogeneous part of
the diffeomorphism group, $\overline{GL}(n, \RR)$ the general linear
group and $\overline{SO}(1,n-1)$ the Lorentz group.  This corresponds
to the breaking of translations (TR), nonlinear transformations (NL),
dilations and shear transformations (SD), respectively.

The existence of such a symmetry breaking scenario appears more convincing
if it is considered from bottom up. At low temperatures the
vacuum is invariant under local Lorentz transformations and matter is
represented by Lorentz spinors.  It is conceivable that at higher
temperatures matter transforms under a larger spacetime group. The
most prominent example is scale invariance which is believed to be
restored at high energies. Here matter is described by spinors of the
conformal group which contains the Lorentz group as a subgroup.  It is
not implausible that further symmetries of the diffeomorphism group
and in the end all of them are restored at very high temperatures.

So far we focused exclusively on the breaking of the transformation
group $H$ of states in the Hilbert space. \mbox{After} a series of
phase transitions, matter is again represented by spinors of the
Lorentz group rather than the diffeomorphism group. The appealing
aspect of this view of matter is that gravitational fields emerge
naturally as Goldstone bosons of the symmetry breaking (quasi as a
by-product).  In each phase of the breaking we loose degrees of
freedom in the matter sector, i.e.\ states in the Hilbert space, but
gain new geometrical objects in terms of Goldstone fields. Spacetime
appears as an emergent product of this process.

A convenient concept to determine these \mbox{Goldstone} bosons is
given by the nonlinear realization approach \mbox{{\cite{Colea,Coleb,
Salaa, Salab}}}. This technique provides the transformation behavior of
fields~$\xi$ of a coset space $G/H$ which is associated with the
spontaneous breaking of a symmetry group $G$ down to a stabilizing
subgroup $H$. Nonlinear realizations of spacetime groups have been
studied in a number of papers \cite{Isha, Bori, Pash, Tseytlin, Tres, Puli,
Lope, Ogie74, Lord1, Lord2, Ivanov:1981wn}.  As in \cite{Bori, Pash}
we consider nonlinear realizations of the diffeomorphism group, i.e.\
we choose $G=\textit{Diff}\,(n, \RR)$ and $H$ to be one of the groups
in the above sequence. It turns out that in this way the relevant
gravitational fields, such as coordinates, affine connection and metric,
are all of the same nature: They can be identified with Goldstone
bosons or coset fields of the diffeomorphism group.

Let us have a brief look at the nonlinear realizations in detail.  The
first nonlinear realization with $H_0=\textit{Diff}_0(n, \RR)$
corresponds to the breaking of translational invariance and shows the
existence of dynamical coordinates in terms of Goldstone fields.
Subsequently, we realize $\textit{Diff}\,(n, \RR)$ with
$H_1=GL(n,\RR)$ as stability group.  The corresponding coset fields
transform as a holonomic affine connection and can be used for an
affine theory of gravity. Finally, we realize $\textit{Diff}\,(n,
\RR)$ with the Lorentz group $H_2=SO(1,n-1)$ as stability group
corresponding to the additional breaking of dilations and shear
transformations. The corresponding coset parameters can be interpreted
as tetrads which lead to the definition of a metric.  This has already
been found by Borisov and Ogievetsky \cite{Bori} who studied a
simultaneous realization of the affine and the conformal group with
the Lorentz group as stability group.

\begin{figure}
\includegraphics[scale=0.7]{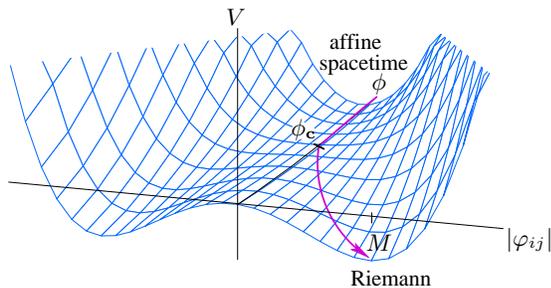}
\caption{The potential $V(\varphi_{ij}, \phi)$. The metric is
conformally flat for $\phi> \phi_c$, tachyonic for $\phi<\phi_c$ and
becomes massless at the minimum of the potential at $\phi=0$, $\vert
\varphi_{ij} \vert = \pm M$.} \label{fig}
\vspace{-5.4cm} \hspace{-1.1cm} $V$ \vspace{4.9cm}

\vspace{-4.4cm} \hspace{2.7cm} $\phi$ \vspace{3.9cm}

\vspace{-2.2cm} \hspace{6.7cm} $\vert \varphi_{ij}\vert$ \vspace{1.7cm}

\vspace{-2cm} \hspace{2.7cm} ${M}$ \vspace{1.5cm}

\vspace{-3.4cm} \hspace{0.7cm} $\mathbf{\phi_c}$ \vspace{3.4cm}
\end{figure}
Though nonlinear realizations of spacetime groups have been studied
for quite some time, with a few exceptions \cite{West, Kirsch, Sija88,
Leclerc:2005qc, Percacci:1990wy, Percacci, Wilczek:1998ea}, there have
not as yet been developed any Higgs models for the dynamical breaking
of these groups.  In this paper we construct such a gravitational
Higgs mechanism by introducing the metric as a Higgs field into an
affine spacetime. This effectively corresponds to a metric-affine
theory of gravity \cite{MAG} in which $GL(4,\RR)$ breaks down to
$SO(1,3)$ in the tangent space of the spacetime manifold.

The Higgs sector is constructed as follows.  In analogy to the
isospinor scalar $\Phi=(\phi^+, \phi^0)$ of electroweak symmetry
breaking, the breaking is induced by a (real) scalar field $\phi$ and
a symmetric tensor~$\varphi_{ij}$ which has ten independent
components. Under the Lorentz group the tensor $\varphi_{ij}$
decomposes into its trace $\sigma$ and a traceless tensor $\hat
\varphi_{ij}$ according to ${\bf 10 \rightarrow 1 + 9}$. The singlet
$\sigma$ turns out to be a massive gravitational Higgs field, whereas
the fields $\hat \varphi_{ij}$ and $\phi$ are the ten Goldstone fields
associated with the coset $GL(4, \RR)/SO(1, 3)$. As shown by the
nonlinear realization approach, these fields define the metric
\begin{align}
g_{ij} =  e_i{}^\alpha e_j{}^\beta \eta_{\alpha\beta} \,,\quad
e_i{}^\alpha \equiv \phi \exp \left( i \hat \varphi_{jk} \hat T^{jk}
\right){}_i{}^\alpha \,,
\end{align}
where $\hat T^{ij}$ are the shear generators and $\phi$ parameterizes
dilations. In Tab.~\ref{tab2} we compare the Higgs sector of the
electroweak symmetry breaking with that in gravity.

\begin{table}[h]
\begin{center}
\begin{tabular}{cccccc}
  \hline {\bf symmetry breaking}& {\bf electroweak} & {\bf
   gravity}\\ \hline symmetry  & $SU(2) \times U(1)_Y$ & $GL(4, \RR)$
   \\ stabilizer & $U(1)_{EM}$ & $SO(1,3)$ \\ Higgs field $\Phi$ & 
$(\phi^0, \phi^+)$
  & $(\varphi_{ij}, \phi)$ \\ \# components of $\Phi$ & $4$ & $10+1$\\
  \# Goldstone bosons & $3$ & $10$ \\
  \# Higgs particles & $1$ & $ 1$ \\
   massive bosons & $W^\pm, Z$ &  $Q_{ijk}$ \\ 
\hline
\end{tabular}
\end{center} \caption{A comparison of the electroweak symmetry breaking
and the Higgs mechanism in gravity. $Q_{ijk}\equiv\nabla_k g_{ij}$ is the
nonmetricity tensor.} \label{tab2}
\end{table}

The symmetry breaking potential $V(\varphi_{ij}, \phi)$ in our model
is similar to that of hybrid inflation \cite{hybrid} with $\phi$
the inflaton and $\varphi_{ij}$ replacing the non-inflaton scalar. As
shown in Fig.~\ref{fig}, the field $\phi$ rolls down the channel at
$\varphi_{ij}=0$ until it reaches a critical value $\phi_c$ at which
point $\varphi_{ij}=0$ becomes unstable and the field rolls down to
the minimum of the potential at $\phi=0$ and $\vert \varphi_{ij} \vert
= \pm M$. In other words, the breaking of dilations triggers the
spontaneous breaking of shear symmetry and induces the condensation of
the metric $g_{ij}$.

During the condensation the affine connection absorbs the metric. Some
degrees of freedom of the connection known as nonmetricity $Q_{ijk}$
acquire a mass as a consequence of ``eating'' the Goldstone
metric. This is the analog of the absorption of Goldstone bosons by
the gauge bosons of $SU(2) \times U(1)_Y$ which become massive $W^\pm$
and~$Z$ bosons. The mass of the nonmetricity is however of order of
the Planck scale such that nonmetricity decouples at low energies. If
we also neglect torsion, then the affine connection turns
into the \mbox{Christoffel} connection and we recover an effective 
Riemannian spacetime at the minimum of the potential, see
Fig.~\ref{fig}.

The paper is organized as follows. In section~\ref{sec2} we review
several aspects of the diffeomorphism algebra and sketch the nonlinear
realization technique in order to fix the notation. We also discuss
principles of gravity and their relation to nonlinear realizations of
the diffeomorphism group.  In section~\ref{sec3} we show that
coordinates, metric and connection can all be identified as Goldstone
bosons of the diffeomorphism group. In section~\ref{sec4} we construct
the Higgs model for the condensation of the metric in an affine
spacetime. We conclude in section~\ref{sec5} with some final remarks
and some open questions. Many detailed computations of the nonlinear
realizations can be found in the appendix.

%%%%%%%%%%%%%%%%%%%%%%%%%%%%%%%%%%%%%%%%%%%%%%%%%%%%%%%%%%%%%%%%%%%%%
\section{Principles of gravity and nonlinear
realizations of the diffeomorphism group} \label{sec2} 
%%%%%%%%%%%%%%%%%%%%%%%%%%%%%%%%%%%%%%%%%%%%%%%%%%%%%%%%%%%%%%%%%%%%%

In this section we briefly review the algebra of analytic
diffeomorphisms and the nonlinear realization technique. We also
discuss the relation between principles of gravity and nonlinear
realizations of the diffeomorphism group. We suggest that the two
groups involved in these nonlinear realizations are unambiguously
fixed by the principle of general covariance and an appropriate
equivalence principle.

\subsection{Principles of theories of gravity}
Classical theories of gravity are explicitly or implicitly based on
two \emph{invariance principles}. The first one is the principle of
general covariance. What is actually meant by general covariance has
often been the subject of discussion in the literature, see e.g.\
Ref.~\cite{9910079}. General covariance does not just mean invariance under
general coordinate transformations, since every theory can be made
invariant under (passive) diffeomorphisms as has already been pointed
out by Kretschmann \cite{Kret}. It is not obvious how the group of
diffeomorphisms selects the metric or the affine connection as the
dynamical field in a theory of gravity. We will show below that
nonlinear realizations of the diffeomorphism group give these fields a
group \mbox{theoretical} foundation.

In addition to the principle of general covariance, theories of
gravity also require a hypothesis about the geometry of spacetime.
The latter is mostly disguised in the formulation of an {equivalence
principle} (EP). We know at least three classical EP's, see
e.g.~\cite{Laem}: the weak (WEP), the strong (SEP) and Einstein's
equivalence principle (EEP).  Each EP determines a particular
geometry: While the WEP postulates a quite general geometry
(Finslerian e.g.), the SEP restricts spacetime to be affine. Finally,
EEP assumes local Lorentz invariance leading to a Riemannian geometry.

From the perspective of the nonlinear realization technique, it is not
a coincidence that theories of gravity are based on exactly two
postulates.  As we will review in Sec.~\ref{secnlr}, there are two
groups, $G$ and $H$, involved in a nonlinear realization: The group
$G$ is represented nonlinearly over one of its subgroups $H$. It is
quite plausible that these groups are fixed by the principle of
general covariance and an appropriate equivalence principle. The
former fixes $G$ to be the diffeomorphism group, $G=\textit{Diff}\,(n,
\RR)$, while the later fixes $H$ to be either the general linear group
$H_1=GL(n, \RR)$ (in case of the SEP) or the Lorentz group
$H_2=SO(1,~n-1)$ (in case of the EEP). We will see in Sec.~\ref{sec3}
that such nonlinear realizations lead to an affine or a Riemannian
spacetime, respectively.

\subsection{The group of analytic diffeomorphisms}
In the following we briefly discuss the algebra of analytic
diffeomorphisms. Representations of the group \textit{Diff}$(n,\RR)$
can be defined in an appropriate Hilbert space of analytic functions
$\Psi_A(x^i)$. We make the assumption that the manifold, on which
$\Psi_A$ is defined, locally allows for a Taylor expansion.  The
generators can be expressed in terms of coordinates $x^i$ as
$(m=-1,...,\infty$; $i, j_a=0,...,n-1$; $a=1,...,m+1$) \cite{Sija83,
Bori2}
\begin{align} \label{rep}
  &F^m_i{}^{j_1...j_{m+1}}\Psi_A= \nonumber\\
 &\hspace{.5cm}\underbrace{i x^{j_1}...x^{j_{m+1}}
    \frac{\partial}{\partial x^i}\Psi_A}_{\rm orbital}
  +\underbrace{i(\hat F_i{}^{j_1...j_{m+1}})_A{}^B \Psi_B \frac{}{}
    }_{\rm intrinsic} \,
\end{align}
with the intrinsic part $(m \geq 0)$
\begin{align}
\hat F_i{}^{j_1...j_m}=\partial_k( x^{j_1}...x^{j_m}) \hat L_i{}^k,
\end{align}
where $\hat L_i{}^k$ are representations of $\overline{GL}(n, \RR)$.  The
generators have one lower index and are symmetric in the $m+1$ upper
indices. The lowest generators ($m=-1,0$) are the translation
operators $P_i\equiv F_i^{-1}$ and the operators of the linear group
$L_i{}^j\equiv F^0_i{}^j$. Generators $F^m$ with $m \geq 1$
generate nonlinear transformations.

The generators of \textit{Diff}$(n,\RR)$ satisfy the commutation relations
\begin{align} \label{alg}
&[F^n_k{}^{i_1...i_{n+1}}, F^m_l{}^{j_1...j_{m+1}}] = \nonumber \\
   &\hspace{.5cm}= i \sum\limits_{a=1}^{m+1} \delta^{j_a}_k 
      F^{m+n}_l{}^{i_1...i_{n+1}j_1...\hat j_a...j_{m+1} } \nonumber\\
   &\hspace{.5cm}- i \sum\limits_{a=1}^{n+1} \delta^{i_a}_l 
      F^{m+n}_k{}^{i_1...\hat i_a...i_{n+1}j_1...j_{m+1} }  ,
\end{align}
where the indices with a hat are omitted. Two important
subalgebras are the algebras of the linear group $GL(n,\RR)$ and the 
Lorentz group $SO(1,n-1)$ with commutation relations 
\begin{align}
[L_i{}^j, L_k{}^l] = i \d_i^l L_k{}^j  
                                   - i \d_k^j L_i{}^l \,
\end{align} and \begin{align}
[M_{ij}, M_{kl}]=i\eta_{il}M_{kj}- i\eta_{jl}M_{ki}
                                -i\eta_{kj}M_{il}+ i\eta_{ki}M_{jl},
\end{align}
where $M_{ij}\equiv L_{[i}{}^k \eta_{j]k}$ are the Lorentz generators.

\subsection{Nonlinear realizations} \label{secnlr}
Let us briefly summarize the nonlinear realization technique
\cite{Colea, Coleb, Salaa, Salab}. In order to fix the notation, we
only list some important formulas which we use throughout this paper.

Nonlinear realizations are based on the notion of a fiber bundle.  Let
$H$ be a closed not invariant subgroup of a Lie group $G$. Then $G/H=
\{gH, g \in G \}$ is a \textit{homogeneous space} not a group and $G$
can be decomposed as
\begin{align}
 G=\{H \cup g_1H \cup g_2 H \cup ...\}  ,
\end{align}
where $g_1 \notin H, g_2 \notin \{H, g_1H\}$, etc. This means that $G$
can be written as a union of spaces $\{ g_i H \}$, all diffeomorphic
to $H$ and parameterized by the \textit{coset space} $G/H$.  So the
group $G$ can be regarded as a principal fiber bundle with structure
group $H$, base space $G/H$, and projection $\sigma^{-1}: G \rightarrow G/H$.

Assume a group $G$ shall be represented nonlinearly over one of its
subgroups $H$. Following \cite{Colea, Coleb, Salaa, Salab}, the
fundamental nonlinear transformation law for elements $\sigma$ of
$G/H$ is given by
\begin{align} \label{A12}
g \sigma(\xi) = \sigma(\xi') h(\xi,g)\, .
\end{align}
An element $\sigma(\xi)$ is transformed into another element
$\sigma(\xi')$ by multiplying it with $g \in G$ from the left and with
$h^{-1} \in H$ from the right.

The standard form of an element $\s \in G/H$ is given by
\begin{align}
\s(\xi)\equiv e^{i \xi^i A_i}\, ,
\end{align}
where $A_i$ are the generators of the coset space $G/H$ and $\xi^i$
the corresponding coset parameters. Eq.\ (\ref{A12}) defines
implicitly the nonlinear transformation $\xi \rightarrow \xi'$, i.e.\
the transformation behavior $\delta \xi$ of the coset parameters
$\xi$.

In the nonlinear realization of symmetry groups the total connection
is given by the Maurer-Cartan 1-form
\begin{align}
\Gamma\equiv \sigma^{-1}d\sigma \,.
\end{align}
By differentiation of Eq.~(\ref{A12}) with respect to the coset
fields~$\xi$, we obtain $gd\s=d\s' h + \s'dh$ and from this the nonlinear
transformation law
 \begin{align} \label{tl}
 \Gamma'= h \Gamma h^{-1} + h d h^{-1} \,.
\end{align}

The total connection $\Gamma$ can be divided into pieces $\G_{\rm H}$
and $\G_{\rm G/H}$ defined on the subgroup $H$ and the space $G/H$,
respectively. The transformation law (\ref{tl}) then shows that
$\G_{\rm H}$ transforms inhomogeneously, whereas $\G_{\rm G/H}$
transforms as a tensor:
\begin{align} 
 \G'_{\rm H}&= h \G_{\rm H} h^{-1} + h d h^{-1}\,, \nonumber\\
 \G'_{\rm G/H}&= h \G_{\rm G/H} h^{-1}\,. \label{A252}
\end{align}
In other words, only $\G_{\rm H}$ is a true connection which can be
used for the definition of a covariant differential
\begin{align}
D \psi \equiv  (d + \Gamma_{\rm H}) \psi \,  
\end{align}
acting on representations $\psi$ of $H$. 

At first sight one might think that the curvature $R\equiv d \Gamma +
\Gamma \wedge \Gamma$ vanishes identically since $\Gamma=\s^{-1}
d\s$. However, any Cartan form with a homogeneous transformation law
can be put equal to zero \cite{Ivan}. This is an invariant condition
and does not affect physics.  Because of the homogeneous transformation
behavior of $\G_{\rm G/H}$, only the curvature $R_{\rm H}\equiv d
\G_{\rm H} + \G_{\rm H} \wedge \G_{\rm H}$ is physically relevant.

%%%%%%%%%%%%%%%%%%%%%%%%%%%%%%%%%%%%%%%%%%%%%%%%%%%%%%%%%%%%%%%%%%%%%
\section{Nonlinear realizations of the diffeomorphism group} \label{sec3}
%%%%%%%%%%%%%%%%%%%%%%%%%%%%%%%%%%%%%%%%%%%%%%%%%%%%%%%%%%%%%%%%%%%%%

In this section we nonlinearly realize the group of diffeo\-morphisms
$G=\textit{Diff}\,\mbox{$(n,\RR)$}$ over the homogeneous part of the
diffeomorphism group $H_0=\textit{Diff}_0\mbox{$(n,\RR)$}$, the
general linear group $H_1=GL(n,\RR)$ and the Lorentz group
\mbox{$H_2=SO(1,n-1)$}. We show that the parameters of the
corresponding coset spaces $G/H$ ($H \in \{H_1,H_2\}$) can be
identified with the geometrical objects of an affine and a Riemannian
spacetime, respectively. In particular, we find that the parameters
$\xi^i$, $h^{ij}$, and $\omega^i{}_{jk}$ associated to the generators
$P_i \equiv F^{-1}_i$ (translations), $T_{ij} \equiv F^0_{(ij)}$
(shear transformations), and $F^1_i{}^{jk}$ transform as coordinates,
metric, and affine connection.

\subsection{The origin of a manifold ($\mathbf{G}$-coordinates)}
\label{sec21}

The basic component of spacetime is a differentiable manifold with a
local coordinate system. In this section we reveal the group
theoretical origin of coordinates by constructing the coset space
$G/H_0$ with $G=\textit{Diff}\,(n, \RR)$ and $H_0=\textit{Diff}_0\,(n,
\RR)$ its homogeneous subgroup. This corresponds to the breaking of
translations $x^i \rightarrow x^i + a^i$ in the representation space
of the diffeomorphism group.

For the construction of the coset $G/H_0$ it is convenient to write an
element $g\in G$ as
\begin{align} \label{g}
g=e^{i \ve^i P_i} e^{i \ve^i{}_j 
  L_i{}^j } e^{i \ve^i{}_{jk} F^1_i{}^{jk}}u \,,
\end{align}
with $u$ para\-meterizing that part of the diffeomorphism group which
is spanned by the generators $F^n$~$(n \geq 2)$.  The coset space
$G/H_0$ is just spanned by the translation generators $P_i$ and
elements $\s \in G/H_0$ can be written as
\begin{align}
 \s=e^{i \x^i P_i} \,,
\end{align}
where the fields $\x^i$ are the corresponding coset parameters.

In general, the total nonlinear connection $\G$ can be expanded in the
generators of $\textit{Diff}\,(n,\RR)$ as
\begin{align} \label{A29}
 \Gamma \equiv \s^{-1}d\s= i \vartheta^i P_i + i \G^i{}_j L_i{}^j + i
                 \G^i{}_{jk} F^1_i{}^{jk} + ... \,,
\end{align}
i.e.\ $\G$ can be divided into a translational $\vartheta^i$, a linear
$\G^i{}_j$ and a nonlinear part, where $\vartheta^i$, $\G_i{}^j$,
etc.\ are vector-, tensor-valued etc.\ 1-forms, respectively.

The coset parameters $\xi^i$ have three interesting properties. First,
as shown in App.~\ref{app1}, the transformation law (\ref{A12}) for
the elements $\sigma$ determines the transformation behavior of $\x^i$
under $\textit{Diff}\,(n, \RR)$,
\begin{align} \label{transxi}
  \delta \x^i = \ve^i(\xi) \equiv \ve^i + \ve^i{}_j \x^j +
  \ve^i{}_{jk} \xi^j \x^k + ... \,.
\end{align}
This is the transformation behavior of {\em coordinates} 
leading to the interpretation of the base space $G/H_0$ as a
differentiable manifold with coordinates $\xi^i$. Second, the
translational piece of the total connection $\G$ turns out to be the
coordinate coframe $\vartheta^i\equiv d\xi^i$ as can be seen by
computing Eq.~(\ref{A29}). Consequently, the coordinates $\xi^i$
cannot be distinguished from ordinary coordinates. Third, as shown by
the computation in App.~\ref{app1}, the parameters of the
diffeomorphism group are promoted to fields which become explicit
functions of the coordinates $\xi^i$, i.e.~$\epsilon^i \rightarrow
\epsilon^i(\xi)$, $\epsilon^i{}_j \rightarrow \epsilon^i{}_j(\xi)$,
etc. This will become important below when we interpret other
parameters of the diffeomorphism group as geometrical fields.

\begin{figure}
\scalebox{0.54}{\input{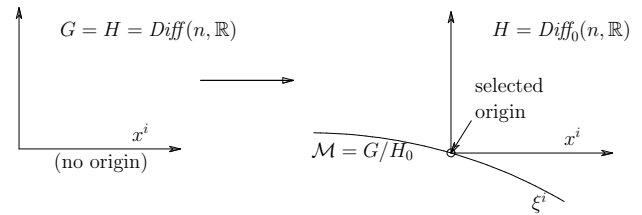}}
\caption{The breaking of translational invariance.} \label{fig2}
\end{figure}

The breaking of global translations $x^i \rightarrow x^i + a^i$ in the
representation space $\cal H$ of $\textit{Diff}\,(n, \RR)$ is achieved
by the selection of a preferred point or an origin in this space. This
is shown in Fig.~\ref{fig2}. In this way no further symmetries are
broken. The origin arises naturally in nonlinear realizations of
$\textit{Diff}\,(n, \RR)$ as the point at which the \mbox{manifold}
${\cal M}=G/H_0$ is attached (``soldered'') to ${\cal H}$. This
reflects the fact that the representation space ${\cal H}$ has become
the tangent space of the manifold $G/H_0$.\footnote{A similar
soldering mechanism has been discussed in \cite{Gronwald} in the
context of Metric-Affine Gravity \cite{MAG}.}

A comment about the difference of the coordinates $x^i$ and $\xi^i$ is
in order. The coordinates $x^i$ are non-dynamical and are necessary
for the definition of the representation~(\ref{rep}) of
\textit{Diff}$(n,\RR)$. These coordinates should not be confused with
the coordinates $\xi^i$.  In contrast to $x^i$, the coordinates
$\xi^i$ represent a {\em dynamical} field. The dynamical character of
the coordinates $\xi^i$ follows from the nonlinear realization
approach: Each coset field, which is not eliminated by the inverse
Higgs effect \cite{Ivan}, is a Goldstone field and as such a dynamical
quantity.

In Sec.~\ref{sec22} and~\ref{sec23} we will identify both the metric
and the affine connection with further parameters of the
diffeomorphism group, i.e.~both fields turn out to be Goldstone
fields, too. Since metric and connection are generally considered as
dynamical quantities, also the {\em \mbox{Goldstone}
coordinates}~$\x^i$ should be regarded in this way.

In order to be a true Goldstone field, the G-coordinates $\xi^i$ must
result from a broken symmetry. Indeed, {\em global} translations $x^i
\rightarrow x^i + a^i$ are broken in the representation space $\cal H$
and the coset parameters~$\xi^i$ are the corresponding Goldstone
fields. At the same time the parameters $\xi^i$ form the base manifold
$G/H_0$ which is interpreted as a spacetime manifold by the
identification of the $\xi^i$ with coordinates. This leads to
reparameterization or diffeomorphism invariance of the spacetime
manifold, see Eq.~(\ref{transxi}). Due to the isomorphism of the
diffeomorphism group with the group of local translations, ${\cal T}
\approx \textit{Diff}\,(n, \RR)$, we gained {\em local} translational
invariance on the (external) spacetime manifold $G/H_0$ at the expense
of loosing {\em global} translational invariance in the (internal)
representation space~$\cal H$.

This implies that general covariance in the sense of diffeomorphism
invariance on the spacetime manifold $G/H_0$ is not broken and
energy-momentum is preserved. Diffeomorphism invariance is however
broken in the representation space ${\cal H}$.

Because of their dynamical behavior, G-coordinates may be visualized as a
{\em ``fluid'' of reference} pervading the universe. Considering
G-coordinates as a continuum, we would interpret the dynamical field
$\xi^i(x)$ as the comoving body frame, whereas the nondynamical
coordinates $x^i$ would be the reference frame.  Such a continuum
mechanical view might remind some of the readers of the concept of an
ether. However, ``ether'' is not an adequate name, since
G-coordinates are not a medium consisting of matter fields. They form
a pure gravitational field just like the metric.

It is clear that a dynamical view of coordinates opens up the
possibility of constructing new cosmological models. For instance,
ghost condensation \cite{Nima} describes the field $\phi \equiv \xi^0$
as a non-diluting cosmological fluid which possibly drives the
accelerating expansion of the universe.  We come back to this issue in
Sec.~\ref{secfluid} in which we discuss some properties of
condensation models for the G-coordinates.

\subsection{The origin of an affine connection} \label{sec22}

The emergence of coordinates as coset parameters of
$\textit{Diff}\,\mbox{$(n,\RR)$}$ is not surprising, since the
diffeomorphism group is the group of general coordinate
transformations. It is however remarkable that other gravitational
fields can also be identified with parameters of the diffeomorphism
group. This will now be shown for the affine connection.

For this purpose let us consider the parameters $\omega^i{}_{jk}$
which together with the G-coordinates $\xi^i$ parameterize the coset
$G/H_1$ with $H_1=GL(n, \RR)$.\footnote{There is also an infinite
number of parameters $\omega^i{}_{jkl}$, $\omega^i{}_{jklm}$, etc.\
associated with the generators $F^n (n \geq 2)$ which will turn out to
be unphysical, see below.} An element of $G/H_1$ is given by
\begin{align}
 \s=\tilde \sigma u= e^{i \x^i P_i} e^{i \omega^i{}_{jk} F^1_i{}^{jk}} u
\end{align}
with $u$ as in Eq.~(\ref{g}). The transformation behavior of
$\omega^i{}_{jk}$ is determined by introducing the infinitesimal group
elements $g \in \textit{Diff}\,(n, \RR)$ and $h \in GL(n, \RR)$
together with $\s$ in the nonlinear transformation law (\ref{A12}).
As shown in detail in App.~\ref{app2}, this gives
\begin{align} \label{A36}
\d \omega^i{}_{jk}= \frac{\partial \varepsilon^i}{\partial \x^l} \omega^l{}_{jk}
                 - \frac{\partial \varepsilon^l}{\partial \x^j} \omega^i{}_{lk}
                 - \frac{\partial \varepsilon^l}{\partial \x^k} \omega^i{}_{jl}
                 + \half \frac{\partial^2 \varepsilon^i}{\partial \x^j 
                 \partial \x^k}\,,
\end{align}
which is the transformation behavior of an affine connection.  Due to
the symmetry in the contravariant indices of the generator
$F^1_i{}^{jk}$, the coset parameters $\omega^i{}_{jk}$ are symmetric
in the indices $j$ and $k$. The connection $\omega^i{}_{jk}$ has only
40 independent components instead of~$64$ (for $n=4$).
 
Moreover, for the total nonlinear connection we find
\begin{align}
 \Gamma &=\s^{-1} d \s = u^{-1} \tilde\s^{-1} d \tilde\s u + u^{-1} du 
                   = \tilde\s^{-1} d \tilde\s + O(F^1) \nonumber\\
                   &= e^{-i \omega^i{}_{jk} F^1_i{}^{jk}}  
                     (i d \x^l P_l ) e^{i \omega^i{}_{jk} F^1_i{}^{jk}}
                       + O(F^1)  \nonumber\\
                   &= i d\x^i P_i + \omega^i{}_{jk} d\x^l 
                      [F^1_i{}^{jk}, P_l] + O(F^1) \nonumber\\
                   &= i \vartheta^i P_i + i 
                      {\G^i{}_j} L_i{}^j + O(F^1)  ,
\end{align} 
with 
\begin{align} \label{holconn}
  \vartheta^i\equiv d\x^i, \quad 
  {\G^i{}_j}\equiv \G^i{}_{jk} d\x^k = -2
  \omega^i{}_{jk} d\x^k \,.
\end{align}

The physical part of the total connection $\G$, which acts on matter
via the covariant derivative, is given by its linear part
\begin{align}
\G_{\rm H_1}\equiv  i 
{\G^i{}_j} L_i{}^j 
\end{align}
whose components $\G^i{}_{jk} = -2 \omega^i{}_{jk}$ transform as an
affine connection under general coordinate transformations. The
translational connection $\vartheta^i=d\xi^i$ is again the coordinate
coframe as shown above.

The elements $u$ contribute only to the unphysical part $\Gamma_{G/H}$
of the total connection $\Gamma$ which does not act on matter.  In
nonlinear realizations of spacetime groups, Goldstone's theorem
(``There is a massless particle for each broken symmetry
generator.'')~applies only in a very restrictive way
\cite{Manohar}. Due to the inverse Higgs effect \cite{Ivan}, some of
the broken generators do not give rise to massless modes. Indeed it
can be shown that the coset parameters $\omega^i_{jkl}$,
$\omega^i{}_{jklm}$, etc.~associated with the generators $F^n$
(\mbox{$n \geq 2$}) do not give rise to any additional Goldstone
bosons. In other words, only a finite number of the infinitely many
coset parameters are Goldstone fields.

Finally, we note that if we had gauged the general linear group, as it
is done in \cite{MAG} and related work, we would have gained a linear
connection, too.  This supports an observation made by Ne'eman
\cite{Neemb}: The group $\textit{Diff}\,(n, \RR)$ being represented
nonlinearly over its $GL(n, \RR)$ subgroup resembles the gauging of
$GL(n, \RR)$. The resulting connection can be used for the
construction of an affine theory of gravity with $GL(n, \RR)$ acting
in the tangent space.

\subsection{The origin of a metric and anholonomic tetrads}
\label{sec23}

In the previous nonlinear realizations with the groups $H_0$ and $H_1$
as stabilizing groups, we identified both coordinates and affine
connection with parameters of the diffeomorphism group. In the same
way, we now show that the metric is related to the coset parameters
associated with shear transformations and dilations. We enlarge the coset
space $G/H_1$ by adding the symmetric generators $T_{ij}\equiv
L_{(ij)}$, where $L_{ij} = L_i{}^k \eta_{jk}$.  This corresponds to
the nonlinear realization of $\textit{Diff}\,(n, \RR)$ over
$H_2=SO(1,n-1)$ which was first considered in~\cite{Bori}.

The elements of the coset space $G/H_2$ can be parameterized as
$\s\equiv e^{i \x^i P_i} e^{i h^{ij} T_{ij}} e^{i \omega^i{}_{jk}
F^1_i{}^{jk}} u$.  Let us also define the tensor $r^i{}_j$ as the
exponential of the field $h^i{}_j$, $r^i{}_j\equiv e^{h^i{}_j}= \d^i_j
+ h^i{}_j + \frac{1}{2} h^i{}_k h^k{}_j+ \cdots$, and
$(r^{-1})_i{}^j\equiv e^{-h_i{}^j}$ its inverse. We raise and lower
the indices of the parameter $h^i{}_j$ by means of the Minkowski
metric $\eta_{ij}=(+--\,...)$ which is given as a natural invariant of
the Lorentz group.

As shown in App.~\ref{app3}, the transformation behavior of the
symmetric tensor $r^i{}_\a$ is given by
\begin{align}         \label{A203}
\d r^{i \a} = \frac{\partial \ve^i}{\partial \x^j} r^{j \a} 
- \tilde \ve_\b{}^\a r^{i \b} 
\end{align}
or,\footnote{ $\frac{\partial \x'{}^i}{\partial \x^j} \approx \d^i_j +
\frac{\partial \ve^i}{\partial \x^j}$ diffeomorphism,
$(\Lambda^{-1})^\b{}_\a \approx \d^\b_\a - \tilde \ve^\b{}_\a$ inverse
Lorentz transformation } in finite form,
\begin{align} \label{A28}
r'{}^i{}_\a=\frac{\partial \x'{}^i}{\partial \x^j} 
r^j{}_\b (\Lambda^{-1})^\b{}_\a , 
\end{align} 
where $\Lambda^\b{}_\a$ is a Lorentz transformation.  The upper index of
$r^i{}_\a$ transforms covariantly while the lower one is a Lorentz
index. The different types of indices have been expected from the
transformation law $\s(\x')=g \s(\x) h^{-1}$, since the coset element
$\s(\x)$ is multiplied by an element $g \in \textit{Diff}\,(n, \RR)$
from the left and an element $h^{-1} \in SO(1,n-1)$ from the
right. Therefore, the parameters $r^i{}_\a$ must transform as a
tetrad.

This leads to the distinction between holonomic ($i, j,...$) and
anholonomic indices ($\a, \b,...$). The tensor $r^i{}_\a$ relates
anholonomic tensors $T^{\a_1...\a_m}_{\b_1...\b_n}$ to holonomic ones
$T^{i_1...i_m}_{j_1...j_n}$ according to
\begin{align}
  T^{\a_1...\a_m}_{\b_1...\b_n}=(r^{-1})_{i_1}{}^{\a_1} \cdots
  (r^{-1})_{i_m}{}^{\a_m} r^{j_1}{}_{\b_1} \cdots r^{j_n}{}_{\b_n}
  T^{i_1...i_m}_{j_1...j_n} \,.
\end{align}
Using (\ref{A203}) one can show \cite{Bori} that the tensors $g_{ij}$
and $g^{ij}$ defined by
\begin{align} \label{metric}
g_{ij}\equiv (r^{-1})_i{}^\a (r^{-1})_j{}^\b \eta_{\a\b}, \qquad
g^{ij}\equiv r^i{}_\a r^j{}_\b \eta^{\a\b}
\end{align} 
transform as the covariant and contravariant metric tensor, respectively.

Due to the different indices of $r^i{}_\a$, one might ask whether
$r^{i\a}$ is symmetric as it is expected from the symmetry of the
tensor $h^{ij}$.  Remember that $h=h(\xi,g) \in H$ depends on $g \in
G$ and the coset parameters $\xi$ of $G/H$. Thus the parameters
$\tilde \ve_{\a\b}$ of the Lorentz subgroup are given in terms of the
coset parameters $\xi^i$ and $r^i{}_\a$ as well as the parameters
$\ve^i, \ve^i{}_j$, etc.\ of an element $g \in G$.  They are
implicitly given by the condition $\d r^{[i \a]}=0$, i.e.\ by the
antisymmetric part of Eq.\ (\ref{A203}),
\begin{align}        
\ve^{[i \vert}{}_j(\xi) \, r^{j \vert \a]} = \tilde \ve_\b{}^{[\a} 
r^{i] \b} \,.
\end{align}
Solving this for $\tilde \ve^{\a\b}$, one obtains \cite{Lope}
\begin{align} 
  \tilde \ve^{\a\b}=\b^{\a\b}-\a^{ij} \tanh \left\{\half \log [
    r_i{}^\a (r^{-1})_j{}^\b ] \right\} \,,
\end{align}
whereby $\a_{ij}$ and $\b_{ij}$ are the symmetric and the
antisymmetric part of $\ve^i{}_j(\xi)$. It is this complicated
dependence on the parameters of $G$ and $G/H$ which guarantees the
symmetry of $r^{i\a}$.

We now derive the Christoffel connection of the \mbox{Riemannian}
spacetime.  In App.~\ref{sec1form}, we have calculated the
coefficients $\vartheta^\a$ and $\G^{\a\b}$ of the expansion
(\ref{A29}) of the total connection $\G$. They
are\footnote{$[r^{-1},dr]^{\a\b} \equiv (r^{-1})_i{}^\a dr^{i \b}
-dr^{i \a} (r^{-1})_i{}^\b$,\\ $\{r^{-1},dr\}^{\a\b} \equiv
(r^{-1})_i{}^\a dr^{i \b} +dr^{i \a} (r^{-1})_i{}^\b$}
\begin{align}
\vartheta^\a &\equiv  (r^{-1})_i{}^\a d\x^i ,\\ \G^{(\a\b)} &\equiv \frac{1}{2}
\{r^{-1},dr \}^{\a\b} - \omega^{\a\b\g} \vartheta_\gamma - \omega^{\b\a\g}
\vartheta_\gamma , \label{r1} \\ \Gamma^{[\a\b]} &\equiv \frac{1}{2}
[r^{-1},dr]^{\a\b} - \omega^{\a\b\gamma} \vartheta_\gamma + \omega^{\b\a\gamma} \vartheta_\gamma
\,\label{r2}.
\end{align} 
We now show that the antisymmetric part $\Gamma^{[\a\b]}$ as given by
Eq.~(\ref{r2}) is identical to the Christoffel connection.  In
accordance with the notation used in \cite{Bori}, we define
\begin{align}
\nabla^\g h^{\a\b}\equiv \half r^{i \g} \{ r^{-1}, \partial_i r \}^{\a\b}\,
\end{align}
which is, due to the identity (\ref{A34}), 
\begin{align}
 \nabla_\g h_{\a\b} = - \frac{1}{2} \partial_k g_{ij}  r^i{}_\a
  r^j{}_\b r^k{}_\g  \,,
\end{align}
nothing but the partial derivative of the metric in an anholonomic
frame.  The covariant derivative of $h^{\a\b}$, also known as {\em
nonmetricity}, is defined by
\begin{align} \label{A40}
D^\g h^{\a\b} \equiv  {\Gamma^{\g(\a\b)}}
              & = \nabla^\g h^{\a\b} - \omega^{\a\b\g} - \omega^{\b\a\g} \nonumber\\
              & = \nabla^\g h^{\a\b} - 3 \omega^{(\g\a\b)}
                + \omega^{\g\a\b} \,.
\end{align}

In order to see why nonmetricity vanishes, we make again use of the
inverse Higgs effect \cite{Ivan}. This ``effect'' is based on the fact
that any Cartan form with a homogeneous transformation law can be put
equal to zero without affecting physics. This applies to the part of
the connection $\G$ which is defined on the coset space $G/H$, see
Eq.\ (\ref{A252}). In particular, we set
\begin{align} \label{hconst}
D^\g h^{\a\b} \equiv \G^{\g(\a\b)}=0 \,.
\end{align}
Solving this for $\omega^{\a\b\g}$, we get
\begin{align} \label{A41}
  \omega^{\a\b\g} = - \nabla^\a h^{\b\g} + 3 \omega^{(\a\b\g)} 
\end{align}
which shows that a part of $\omega^{\a\b\g}$ can be expressed by the
Goldstone fields $h^{\a\b}$. We substitute this into $\G^{[\a\b]}$ and
obtain
\begin{align}
 \G^{[\a\b]} & =\frac{1}{2} [r^{-1},dr]^{\a\b} 
               + \nabla^\a h^{\b\g}\vartheta_{\g} - \nabla^\b h^{\a\g}\vartheta_{\g}  
               \nonumber\\
               &+\left(T^{\b\g\a}-T^{\g\a\b}+T^{\a\b\g}\right) \vartheta_\g 
\end{align}
with vanishing torsion
\begin{align}
 T^{\a\b\g}\equiv \G^{\a[\b\g]}=-2 \omega^{\a[\b\g]}=0 \,.
\end{align}
Recall that $\omega^{\a\b\g}$ is symmetric in the last two indices.

This can also be written in components, $\G_{\a\b}=\G_{\a\b
  k} d\xi^k$,
\begin{align} \label{A250}
  \G_{\a\b k}= \half  [r^{-1}, \partial_k r]_{\a\b} - 
  2 (r^{-1})_k{}^\g \nabla_{[\b} h_{\a]\g} \,.
\end{align}

Let us finally calculate the holonomic version of the connection
(\ref{A250}) by means of
\begin{align}
  \G_{ijk}=(r^{-1})_i{}^\a \Gamma_{\a\b k} (r^{-1})_j{}^\b + (r^{-1})_i{}^\a
  \partial_k (r^{-1})_{j\a} .
\end{align}
and the identity (\ref{A34}).  We obtain the Christoffel connection
\begin{align}
\G_{ijk}=\half (\partial_k g_{ij} + \partial_j g_{ik}-\partial_i g_{jk}) 
\end{align} 
of general relativity. We see that the inverse Higgs effect corresponds
to the absorption of the metric in the connection.

To summarize the above realizations, we found that {\em coordinates,
metric} and {\em affine connection}, which appear so differently as
far as their transformation behavior is concerned, are all of the same
nature: They are Goldstone bosons parameterizing coset spaces formed
by the diffeomorphism group and an appropriate subgroup.  The
existence of these geometrical objects has thus been proven by group
theory. 

Let us compare our approach with nonlinear realizations of {\em local}
spacetime groups such as the local Poincar\'e or affine group
\cite{Tseytlin, Tres, Puli}. In these realizations the total nonlinear
connection is given by $\Gamma= \sigma^{-1} (d+ \omega) \sigma$, and
$\omega$ is the linear gauge connection of the local
spacetime group instead of a Goldstone field. In contrast, we
restricted to the nonlinear realization of the diffeomorphism group
which is isomorphic to the group of local translations and as such a
subgroup of the above mentioned local groups. Since, with the
exception of torsion, all relevant spacetime structures could be
obtained by the nonlinear realization technique, it seems to be
sufficient to require only diffeomorphism invariance.

Finally, we would like to mention that it is also possible to recover
torsion within our approach.  If we allowed for non-commutative coordinates
satisfying, for instance,
\begin{align}
[x^i,x^j]=i \theta^{ij} \,,
\end{align}
where $\theta^{ij}$ is a constant antisymmetric tensor of dimension
$-2$, the generators $F^m_i{}^{jk...}$ of the Ogievetsky algebra
would not be symmetric in the upper indices anymore.\footnote{A
$\theta$-deformed algebra of diffeomorphisms has recently been studied
in \cite{Aschieri:2005yw}.} Then the connection $\omega^i{}_{jk}$ would
not be symmetric in the last two indices either and torsion would be
an additional gravitational field. Note, however, that unlike in the gauge
approach to gravity, the existence of torsion is directly linked to the
non-commutativity of coordinates.

%%%%%%%%%%%%%%%%%%%%%%%%%%%%%%%%%%%%%%%%%%%%%%%%%%%%%%%%%%%%%%%%%%%%%
\section{A Higgs mechanism for gravity} \label{sec4}
%%%%%%%%%%%%%%%%%%%%%%%%%%%%%%%%%%%%%%%%%%%%%%%%%%%%%%%%%%%%%%%%%%%%%

In the previous section we discussed nonlinear realizations of the
diffeomorphism group which provide gravitational fields as Goldstone
bosons of a dynamical breaking of $\textit{Diff}\,(n, \RR)$.  In the
following we develop a concrete model for this symmetry breaking.  We
construct a Higgs mechanism which breaks the general linear group
$GL(n,\RR)$ down to the Lorentz group $SO(1,n-1)$. The breaking is
induced by the condensation of the metric which transforms an affine
spacetime into a Riemannian one. We finally comment on the breaking of
global translations which gives rise to the condensation of a fluid of
reference in terms of Goldstone coordinates.

\subsection{Breaking of $GL(n,\RR)$ and hybrid inflation}

Previous Higgs models of the (special) linear group have been
constructed in \cite{Sija88, Kirsch}. In \cite{Sija88} the metric was
independent from the symmetry breaking Higgs fields. This approach is
however not in the spirit of the metric as a Goldstone field as
indicated in \cite{Kirsch}. In the following we construct a Higgs
mechanism in which the degrees of freedom of the metric are identical
to the symmetry breaking fields.

We assume that the stability group of the diffeomorphism group is
$H_1=GL(n,\RR)$ at high energies. According to the nonlinear
realization considered in Sec.~\ref{sec22}, this corresponds to an
affine spacetime which is equipped with an affine connection
$\Gamma^i{}_{jk}$ which is symmetric in the lower indices and has $40$
independent components. For the moment we ignore the dynamics of the
G-coordinates $\xi^i$ and work in the gauge $x^i=\xi^i$.

For the symmetry breaking we also have to introduce a ten-component
second rank symmetric tensor $\varphi_{ij}$ of ${GL}(n,\RR)$ and a
real scalar field~$\phi$.  These fields are the analogs of the
isospinor scalar field $\Phi=(\phi^+, \phi^0)$ which induces the
breaking of the electroweak interaction in the standard model of
elementary particle physics. The tensor $\varphi_{ij}$ decomposes
under the Lorentz group into its trace $\varphi^{(0,0)}$ and a
traceless symmetric tensor $\varphi^{(1,1)}$, i.e. ${\bf 10}
\rightarrow {\bf 9} + {\bf 1}$ for $n=4$.\footnote{The indices
indicate the representation of the Lorentz group labeled by ($j_1,
j_2$).}  The fields $\varphi^{(1,1)}$ and $\phi$ parameterize the
coset $GL(n,\RR)/SO(1,n-1)$, i.e.\ they are \mbox{Goldstone} fields
associated with shears and dilations. The field $\varphi^{(0,0)}$ will
become a massive Higgs field.

We also make use of metric- and tetrad-type fields $g^{ij}$ and
$e_\a{}^i$ which we define in terms of the Goldstone fields
$\varphi^{(1,1)}$ and $\phi$ by
\begin{align}
e_\a{}^i &\equiv \phi \exp \left( i \varphi_{jk}^{(1,1)} \hat T^{jk}
\right){}_\a{}^i \,, \label{tet}\\ g^{ij} &\equiv \eta^{\a\b}
e_\a{}^i e_\b{}^j \,, \label{met}
\end{align}
where $\hat T^{ij}$ and $D=-i$ (in the redefinition $\phi=\exp(i\tilde
\phi D)$) are the generators of shears and dilations, respectively.
The metric $g_{ij}$ may be used for raising and lowering indices.  We
stress that $g_{ij}$ is a descendant of $\varphi^{(1,1)}$ and $\phi$
and not an independent field. This reflects the fact that the metric
is the Higgs field breaking $GL(n,\RR)$ to $SO(1,n-1)$ as predicted in
the above nonlinear realizations, see also \cite{Sardanashvily} in
this context.

We can now write down a $GL(n ,\RR)$ invariant action for the fields
$\G^i{}_{jk}(x)$, $\phi(x)$, $\varphi_{ij}(x)$ and their descendants
$g^{ij}(x)$ and $e_\a{}^i(x)$. It is convenient to split the action
$S$ into three parts,
\begin{align}
S = S_{\rm grav} + S_{\rm SB} + S_{\rm matter} \,,
\end{align}
i.e.\ into a gravitational, a symmetry breaking and a matter action.

The first part $S_{\rm grav}$ describes the nonminimal coupling of the
fields $\phi$ and $\varphi_{ij}$ to gravity in an affine spacetime.
We choose the gravity action
\begin{align}   \label{lgrav}
  S_{\rm grav}= \int d^4 x \sqrt{-g}  
  \left(\frac{\phi^2}{8\omega}- \frac{\xi}{2}\varphi^{ij} \varphi_{ij}
  \right) (  R + {\cal L}_H ) \,,
\end{align}
where $R=g^{ij} {\rm Ric}_{ij}$ is obtained by contracting the affine
Ricci tensor ${\rm Ric}_{ij}=R^k{}_{ikj}(\Gamma, \partial\Gamma)$
with the metric~$g^{ij}$. ${\cal L}_H$ denotes possible
higher order curvature terms.  The dimensionless coupling constants
$\xi$ and $\omega$ guarantee scale invariance on the classical level.

The metric $g^{ij}$ and the connection $\G^i{}_{jk}$ are independent
fields at high energies and the curvature
\begin{align}
R^i{}_{jkl}=\partial_k \Gamma^i_{jl} - \partial_l \Gamma^i_{jk}
+ \Gamma^m_{jl} \Gamma_{mk}^i - \Gamma^m_{jk} \Gamma_{ml}^i
\end{align}
does not depend on $g^{ij}$. Upon writing the connection as a one-form
$\Gamma^i{}_j= \Gamma^i{}_{jk} dx^k$, it transforms as
\begin{align}
\Gamma'{}^i{}_j =  e^i{}_k \Gamma^k{}_l e^l{}_j + 
   e^i{}_k d e^k{}_j \,
\end{align}
under $GL(n, \RR)$, where $e^i{}_j= \exp (i L_\alpha{}^\beta
\psi^\alpha{}_\beta)^i{}_j$ with $L_\alpha{}^\beta$ the generators of
the linear group.
   
The Palatini approach to general relativity tells us that in the
vacuum the curvature scalar in (\ref{lgrav}) alone does not describe
the dynamics of the post-Riemannian pieces of the connection. For
these pieces, we have to add higher order curvature terms like
\begin{align}
{\cal L}_H \sim  {R}_{(ij)} \wedge {}^\star {R}^{ij} \,, 
\end{align}
for instance, where $R_{ij}=R_{ijkl} dx^k \wedge dx^l$ is the
curvature two-form. Since we introduced a metric into an affine
spacetime, gravity can in principle be described by the Metric-Affine
Theory of Gravity (MAG) \cite{MAG}, where further higher order terms
can be found.\footnote{The main difference to MAG is that in the
present condensation model, the metric is a Higgs field and as such
tachyonic at high energies. Moreover, the tetrads are given by Eq.~(\ref{tet}) 
and do not represent an independent field.}

The second part $S_{\rm SB}$ of the action describes the symmetry
breaking to the Lorentz group and is given by
\begin{align} 
S_{\rm SB}
  =\int d^4 x \sqrt{-g} \left[
 \frac{1}{2} g^{ij} D_i \varphi^{kl} D_j \varphi_{kl}
+ \frac{1}{2} g^{ij} \partial_i \phi \partial_j \phi
-V  \right] \, \label{SSB}
\end{align}
with effective potential 
\begin{align}
 V(\phi, \varphi_{ij}) = \frac{\lambda}{4} \left( \varphi^{ij} \varphi_{ij}
 - M^2 \right)^2 + \frac{1}{2} m^2 \phi^2 + \frac{1}{2}
\lambda' \varphi^{ij} \varphi_{ij} \phi^2. 
\end{align}
The covariant derivative on $\varphi_{jk}$ is defined by
\begin{align}
D_i \varphi_{jk} = \partial _i \varphi_{jk} + i \Gamma_{i\a}{}^\beta
(L^\a{}_\beta)_{jk}{}^{mn} \varphi_{mn} \,,
\end{align}
with $(L^\a{}_\beta)_{jk}{}^{mn}$ the tensor representation of $GL(n,
\RR)$.

Let us consider the action $S_{\rm SB}$ in detail. The first two terms
are kinetic terms for the fields $\varphi_{ij}$ and $\phi$. The
quartic term in the potential is the self-interaction of
$\varphi_{ij}$. The effective mass squared of $\varphi_{ij}$ is
$m^2_\varphi = - \lambda M^2 +\lambda' \phi^2$. The scaling dimensions of
the fields $\varphi_{ij}$ and $\phi$ are $[\varphi^{ij}]=[\phi]=1$ and
$\lambda$ and $\lambda'$ are positive dimensionless coupling constants. We
assume $m$ to be small such that the action is classically invariant
under global $GL(n,\RR)$ transformations at high energies. Note
however that scale invariance is softly broken at energy scales of the
order of $\lambda M^2$.

The reader may have noticed the similarity of the {action} $S_{\rm
SB}$ and {\em hybrid inflation} \cite{hybrid}. Instead of two scalars,
the potential in our model depends on a scalar and a second-rank
tensor. While the inflaton $\phi$ remains a scalar, we replaced the
non-inflaton $\s$ by the second-rank tensor $\varphi_{ij}$.  If we
identify the non-inflaton $\s$ with the trace of $\varphi_{ij}$, i.e.\
$\s \equiv \varphi^{(0,0)}$, we recognize the standard hybrid
inflation potential inside the action $S_{\rm SB}$. Another difference
to hybrid inflation is that gravity is not described by general
relativity in our model. Spacetime is affine during the breaking and
becomes Riemannian only at the end of the condensation.

As an aside we remark that matter in an affine spacetime is described
by a spinorial infinite-component field $\Psi$ of
$\overline{GL}(n,\RR)$. It has been suggested \cite{MAG,
Ne'eman:1998uc, Kirsch, Mick} that such a spinor could be described by
an affine extension of the Dirac equation which would follow from the
action
\begin{align}     \label{lmatter}
  S_{\rm matter} =\, & \int d^4 x \sqrt{-g}
  \bar \Psi \eta^{\a\b} X_\a e_\b{}^i D_i \Psi \,  ,
\end{align}
where the generalized Dirac matrices $X_\a$ form a vector operator of
$\overline{GL}(n,\RR)$. The covariant derivative is given by
$D_i=\partial_i + i \G_{i\a}{}^\b (L^\a{}_\b)$ with $L^\a{}_\b$ an
appropriate spinorial representation of $\overline{GL}(n,\RR)$. After
the symmetry breaking to the Lorentz group, the spinor $\Psi$ splits
into a sum of Lorentz representations with an ordinary Dirac spinor as
the lowest component and (\ref{lmatter}) reduces effectively to the
usual Dirac action. Some progress towards such an equation has
recently been made in \cite{Sijacki:2004cb} in which the matrix $X_\a$
has been constructed for a three-dimensional Dirac-like equation.

\subsection{The condensation of the metric}
We now consider the condensation of the metric and connected with it
the rearrangement of the metric into the connection. Recall that
before the condensation the metric $g^{ij}$ and the
connection $\Gamma^i{}_{jk}$ were independent objects. After the
condensation the affine connection turns into the Levi-Civita
connection as already shown in Sec.~\ref{sec23}.

As in hybrid inflation, we assume that the dilaton field $\phi$ is
slow-rolling and large at the beginning of the breaking. The effective
potential $V$ has a minimum at
\begin{align}
v_\varphi \equiv \sqrt {\langle \varphi^{ij} \varphi_{ij} \rangle}
= \sqrt{ M^2 - \frac{\lambda'}{\lambda}\phi^2 }  \,.
\end{align}
As long as the dilaton $\phi$ is larger than the critical value
$\phi^2_c= \lambda M^2/ \lambda'$, the field $\varphi_{ij}$ is trapped at
$\varphi_{ij}=0$.

The condensation starts as soon as the value of $\phi$ falls below
$\phi_c$, $\phi < \phi_c$, at which point the vacuum becomes
meta-stable. Then the field $\varphi_{ij}$ is not trapped at
$\varphi_{ij}=0$ anymore. Due to quantum fluctuations $\varphi_{ij}$
leaves $\varphi_{ij}=0$ and rolls down the ``waterfall'' to its
minimum $v_\varphi = \pm M$ at $\phi=0$. This has been shown in
Fig.~\ref{fig}.

The metric $g^{ij}$ as defined in Eq.~(\ref{met}) has an interesting
behavior during the symmetry breaking. For $\phi > \phi_c$ the metric
is conformally flat, $g^{ij}=\phi^{2} \eta^{ij}$, and the theory is
approximately scale invariant. Below $\phi=\phi_c$ the effective mass
squared $m_\varphi^2$ of $\varphi_{ij}$ gets negative and the metric
becomes tachyonic. This softly breaks scale invariance and induces the
spontaneous breakdown of shear invariance. Finally, at the end of the
condensation, the metric becomes massless.

In order to show that the metric becomes massless at the minimum of
the potential, we parameterize $\varphi_{ij}$ and $\phi$ around the
minimum $v_\varphi$ as\footnote{This is the analog of the
parameterization of the isospinor $\Phi$ in electroweak symmetry
breaking given by
\begin{align}
\Phi =\frac{1}{\sqrt 2} e^{i \mathbf{\omega} \cdot \mathbf{\tau}/v } \left( 
\begin{matrix} 0 \\ v+h \end{matrix} \right)
\end{align}
with $h$ the Higgs field, $\mathbf{\omega}$ the three Goldstone bosons
and $v$ the vacuum expectation value.}
\begin{align}
  \varphi_{ij} &= (v_\varphi + \s ) \eta_{\a\b} \hat r^\a{}_i
  \hat r^\b{}_j \,, \label{split}\\
 \hat r_\a{}^i &= \exp
  \left(\textstyle\frac{i}{v_\varphi} \varphi_{jk}^{(1,1)} \hat T^{jk}
  \right){}_\a{}^i \,,\\
  \phi &= v_\varphi   \exp (\textstyle\frac{i}{v_\varphi} \tilde 
  \phi D) \,,
\end{align}
where the hat denotes traceless tensors and $\sigma \equiv
\varphi^{(0,0)}$. Eq.~(\ref{split}) explicitly expresses the ${\bf 10}
\rightarrow {\bf 9} + {\bf 1}$ decomposition of $\varphi_{ij}$ under
the Lorentz group. In terms of these fields, the tetrad (\ref{tet})
and the metric (\ref{met}) become
\begin{align} 
e_\a{}^i &= v_\varphi 
 \exp (\textstyle\frac{i}{v_\varphi} \tilde \phi D)  \hat r_\a{}^i \,,
\, \\
g^{ij} &= \eta^{\a\b} e_\a{}^i e_\b{}^j \,. \label{redefmetric}
\end{align}                      
Substituting this into the action $S_{\rm SB}$ as given
by Eq.~(\ref{SSB}), we obtain
\begin{align}
S_{\rm SB} =\,\int d^4x &\sqrt{-g} \left[
\frac{1}{2} g^{ij} \partial_i \s \partial_j \s
-\frac{\lambda}{2} M^2 {\s}^2 \right. \nonumber\\
& \left. +\frac{1}{2} v_\varphi^2 
(e_{\b(j} \partial_i e^\b{}_{k)} -\G_{i(jk)})^2  
+ ... \right] \,, \label{sl}
\end{align}
where dots denote mixed and constant terms. The kinetic term for the
field $\sigma$ and the two terms in the second line of (\ref{sl})
originate from the kinetic terms in the action~(\ref{SSB}). The
Goldstone fields $\phi$ and $\varphi^{(1,1)}$ have become massless,
whereas the field $\sigma$ turned into a massive gravitational Higgs
field with mass
\begin{align}
m^2(\s) = \lambda M^2 \,.
\end{align}

Upon redefining the connection
\begin{align} \label{redef}
 \G'_{ijk}= e_{\b j} \partial_i e^\b{}_{k} - \G_{ijk}\,,
\end{align}
the kinetic terms for the Goldstone bosons $\phi$ and
$\varphi^{(1,1)}$ are absorbed by the connection. In other words, the
Goldstone metric $g_{ij}$, which is composed out of these fields, is
``eaten'' by the connection. In App.~C we show that the total number of 
on-shell degrees of freedom is
preserved during this process.

The absorption of the metric turns the symmetric part $Q_{ijk} \equiv
2 \G'_{i(jk)} = \partial_i g_{jk} - 2 \G_{i(jk)}$ into a tensor called
nonmetricity which is the covariant derivative of the metric.  We
find that the nonmetricity $Q_{ijk}$ gets the mass
\begin{align} \label{mass}
m^2(Q_{ijk}) = v_\varphi^2 = M^2 \,.
\end{align}
This is quite analogous to the breaking of the electroweak interaction
in which the $W$ bosons become massive by the absorption of Goldstone
bosons. The kinetic terms of the Goldstone fields turn into mass terms
for the gauge bosons. As we will see below, the vacuum expectation
value $v_\varphi$ is of the order of the Planck scale such that the
nonmetricity $Q_{ijk}$ decouples from the theory.

Moreover, the antisymmetric part $\G'_{i[jk]}$ remains massless and
can be expressed in terms of the {\em condensed} metric $g^{ij}$
given by Eq.~(\ref{redefmetric}). Note that, since the nonmetricity
\mbox{$Q_{ijk}$} is effectively absent at energies far below
$v_\varphi$, we may set \mbox{$\G'_{i(jk)}=0$}.  We showed in
Sec.~\ref{sec23} that for $\G'_{i(jk)}=0$, the antisymmetric part
$\G'_{i[jk]}$ leads to the Christoffel connection. In this
context compare also the redefinition~(\ref{redef}) with Eqs.\
(\ref{r1}) and (\ref{r2}).\footnote{It is a well-known result that if
nonmetricity and torsion are absent, the connection is necessarily the
Christoffel connection.}

The decoupling of nonmetricity at low energies implies that gravity is
effectively described by general relativity at the minimum of the
potential ($\phi=0$, $v_\varphi=\pm M$). Indeed, the action $S_{\rm
grav}$ given by Eq.~(\ref{lgrav}) reduces to
\begin{align}
  {\cal S}_{\rm grav}= - \int d^4 x \sqrt{-g} \frac{\xi}{2} v_\varphi^2 R \,,
\end{align}
where the curvature $R$ is now determined by the Christoffel
connection and thus in terms of the condensed metric. Comparison
with the standard Einstein-Hilbert action $S_{EH}=-M^2_{Pl} \int d^4x
\sqrt{-g} R$ gives a relation between the Planck mass $M_{Pl}$ and the
mass parameter $M$,
\begin{align}
  M_{Pl}^2 = \frac{\xi}{2}  M^2 \,.
\end{align}
Assuming that coupling constants $\lambda$ and $\xi$ are of order
$O(1)$, we see that the Higgs mass and the mass (\ref{mass}) of the 
nonmetricity are of the order of the Planck scale.

%%%%%%%%%%%%%%%%%%%%%%%%%%%%%%%%%%%%%%%%%%%%%%%%%%%%%%%%%%%%%%%%%%%%%
\subsection{Fluid of reference and ghost condensation} \label{secfluid}
%%%%%%%%%%%%%%%%%%%%%%%%%%%%%%%%%%%%%%%%%%%%%%%%%%%%%%%%%%%%%%%%%%%%%

The breaking of shears and dilations is only one part of the dynamical
breaking of the diffeomorphism group. In a symmetry breaking scenario
of the entire group, one would have to add another mechanism for the
breaking of global translations.  As discussed in Sec.~\ref{sec21},
this corresponds to the construction of a dynamical model for
Goldstone coordinates. In the following we briefly comment on
properties of such a model.

A dynamical model for Goldstone coordinates is in the line of
\cite{Nima, Arkani-Hamed:2003uz} (for early work see \cite{Green,
Siegel}). In these papers the breaking of time-translations leads to
the introduction of a Goldstone scalar $\phi \equiv \x^0$ with a
negative kinetic term.  It seems natural to also break spatial
translations and to introduce kinetic terms for the spatial
coordinates $\x^a$ ($a=1,2,3$), see \cite{Dubovsky, Gripaios:2004ms}
for recent developments.

In these models the prototype for the kinetic term of the field
$\x^i$ ($i=0,1,2,3$) is given by the nonlinear sigma model
\begin{align} \label{sigmamodel}
{\cal S}_\x &= \int d^4 x \sqrt{h} h^{\a\b}(x)
\partial_\a \x^i \partial_\b \x^j g_{ij}(\x) \,.
\end{align} 
This is a classical action for a four-dimensional world volume with
world volume metric $h_{\a\b}$.  The coordinates $x^\a$
parameterize the world volume, while $\x^i$ are the Goldstone bosons
corresponding to the breaking of time and spatial translations in field space.

In the above models $h_{\a\b}$ is interpreted as the spacetime metric
and $g_{ij}$ as some internal metric. Choosing the metric
$g_{ij}=\eta_{ij}$, the ``wrong-sign'' (negative) kinetic term of the
{ghost} field $\phi \equiv \x^0$ in \cite{Nima} follows automatically
from the signature $(- + + +)$ of the Minkowski metric $\eta_{ij}$
($\xi^a=0$ there). In this paper, the nonlinear realizations of
Sec.~III show that the spacetime metric is predominantly a function
of the dynamical field $\x^i$ and one would interpret $g_{ij}$ in
Eq.~(\ref{sigmamodel}) as the target space metric and $h^{\a\b}$ as
the metric in the tangent space.

It is a well-known result from string theory that in flat space
actions of the type (\ref{sigmamodel}) are quantum-mechanically
well-defined only in ten dimensions. However, even on the classical
level, Lorentz-invariant actions for Goldstone coordinates usually
suffer from the van Dam-Veltman-Zakharov (vDVZ) discontinuity
\cite{vanDam:1970vg, Zakharov} or become strongly coupled at very low
energies \cite{Arkani-Hamed:2002sp}.\footnote{Note that in unitary
gauge, actions of the type (\ref{sigmamodel}) lead to the Fierz-Pauli
theory of massive gravity. A more sophisticated Lagrangian than
(\ref{sigmamodel}) is given by \cite{Arkani-Hamed:2002sp}
\begin{align}
{\cal L} &= \sqrt{h} h^{\a\b}(h_{\a\b}-G_{\a\b}) g^{\rho\sigma}
(h_{\rho\sigma}-G_{\rho\sigma}) \,,
\nonumber
\end{align}
with $G_{\a\b} = \partial_\a \x^i \partial_\b \x^j g_{ij}(\x)$. In unitary
gauge ($\xi^i = \Lambda^2 x^i$, $g_{ij}=\eta_{ij} + h_{ij}$), this
leads to the Fierz-Pauli term $h^{\a\b} h_{\a\b} - h^2$.}

In conclusion, it is remarkably difficult to find a proper Lorentz
invariant action for the Goldstone coordinates $\xi^i$. This is also
related to the fact that the $\xi^i$ do not transform as an
irreducible representation under the Lorentz group. Note that the
vector representation of $GL(4, \RR)$ decomposes as ${\bf 4
\rightarrow 1+1+2}$ under the Lorentz group.  For this reason current
models \cite{Nima, Arkani-Hamed:2003uz, Dubovsky} give up the
requirement of Lorentz invariance. It can be shown that under certain
requirements Lorentz-violating actions are free of strong coupling
problems and the vDVZ discontinuity is absent.

%%%%%%%%%%%%%%%%%%%%%%%%%%%%%%%%%%%%%%%%%%%%%%%%%%%%%%%%%%%%%%%%%%%%% 
\section{Conclusions} \label{sec5} 
%%%%%%%%%%%%%%%%%%%%%%%%%%%%%%%%%%%%%%%%%%%%%%%%%%%%%%%%%%%%%%%%%%%%%

We elaborated on the idea that the evolution of the early universe
started with a series of phase transitions in which the Riemannian
spacetime arose step-by-step out of spacetimes with less structure.
We gave some evidence for this view of spacetime by considering
nonlinear realizations of the diffeomorphism group which determine the
field content of gravity in each phase of the symmetry breaking. This
lead to a unified description of several gravitational fields in terms
of Goldstone fields.  A~summary of the broken generators and the
corresponding Goldstone fields is given in Tab.~\ref{tab}.

\begin{table}[ht]
\begin{center}
\begin{tabular}{cccccc}
  \hline {\bf broken symmetry} & {\bf geometrical field} & {\bf
   spacetime}\\ \hline translations $P_i$ &Goldstone- & differential\\
   & coordinates $\xi^i$ & manifold\\ nonlinear $F^1_i{}^{jk}$ &
   connection $\Gamma^i{}_{jk}$ & affine\\ shears/dilations $T_{ij}$ &
   metric $g_{ij}$ & Riemannian\\ \hline
\end{tabular}
\end{center} \caption{Goldstone fields provided by nonlinear realizations
of $\textit{Diff}\,(n, \RR)$. } \label{tab}
\end{table}

We also constructed a Higgs mechanism for the breaking $GL(4, \RR)
\rightarrow SO(1,3)$ in which a Riemannian spacetime emerged out of an
affine spacetime by the condensation of a metric.  The symmetry
breaking potential was very similar to the potential of hybrid
inflation. However, ordinary inflation scenarios assume the validness
of general relativity during inflation. In our model Einstein's theory
is a good description of gravity only at the very end of the
condensation. Nevertheless, it is very suggestive to consider hybrid
inflation as a consequence of the symmetry breaking. In order to show
this, one would have to derive a Friedmann equation for a
(metric-)affine spacetime. The existence of such an equation has
already been shown for a Cartan-Weyl spacetime \cite{Puetzfeld,
Puetzfeld2, Babourova:2002fn}. This gives some confidence that hybrid
inflation could indeed be an artifact of the condensation of the
metric.

We believe that the story of an emergent spacetime has just begun. In
this paper we focused mainly on the breaking of shears and dilations
and only briefly sketched other parts of the full gravitational
symmetry breaking scenario. In particular, there is the condensation
of the \mbox{Goldstone} coordinates corresponding to the breaking of
translational invariance as described in ghost \mbox{condensation} models. It
would be interesting to combine such models with the condensation of
the metric.  It is also conceivable that the affine spacetime itself
arose out of a spacetime with even less structure by the condensation
of an affine connection. We leave this for future research.

\bigskip

\section*{Acknowledgments} 

I would like to thank N.~Arkani-Hamed, F.~W.~Hehl, A.~Miemiec,
L.~Motl, A.~Nicolis and J.~Thaler for many useful
discussions related to this work.  I also would like to thank
R.~Tresguerres and Dj.~$\rm{\check{S}ija\check{c}ki}$ for helpful
comments in the very early stages of this project. This work was
supported by a fellowship within the Postdoc-Programme of the German
Academic Exchange Service (DAAD), grant D/04/23739.

\appendix
%%%%%%%%%%%%%%%%%%%%%%%%%%%%%%%%%%%%%%%%%%%%%%%%%%%%%%%%%%%%%%%%%%%%%
\section{Transformation behavior of coset fields} \label{Ap}
%%%%%%%%%%%%%%%%%%%%%%%%%%%%%%%%%%%%%%%%%%%%%%%%%%%%%%%%%%%%%%%%%%%%%

In this appendix we derive the transformation behavior of the coset
fields $\xi^i$, $r^i{}_\alpha$ and $\omega^i{}_{jk}$ by considering
nonlinear realizations of the diffeomorphism group.  We will repeatedly
use the Campbell-Baker-Hausdorff formula
\begin{align}
e^A B  e^{-A}=B+[A,B]+\frac{1}{2!}[A,[A,B]]+ ...\,
\end{align}
for two matrices $A$ and $B$.

\subsection{The transformation behavior of the coordinates~$\x^i$} 
\label{app1}

In the first nonlinear realization of $\textit{Diff}\,(n, \RR)$, we
choose $H_0=\textit{Diff}_0(n, \RR)$ such that the coset space $G/H$
is only spanned by the translation generators $P_i$. We parameterize
$\s \in G/H, g \in G$, and $h \in H$ as
\begin{align}
\s =& e^{i \x^m P_m }  , \label{seqn}\\
g \approx& 1 + i \ve^i P_i + i \ve^i{}_j L_i{}^j 
             + i \ve^i{}_{jk} F^1_i{}^{jk} + ... , \label{geqn}\\
h \approx& 1 + i \tilde \ve^i{}_j L_i{}^j + ... \label{heqn}\, .
\end{align}
Eq.\ (\ref{A12}), $g \s(\xi) = \s(\xi') h(\xi,g)$, describes implicitly the
transformation behavior $\d\x^i$.  Substituting
Eqns.~(\ref{seqn})--(\ref{heqn}) into (\ref{A12}), we find
\begin{align} \label{sghs}
&e^{-i\x^m P_m} (1 + i \ve^i P_i + ...) e^{i\x^m P_m} = \nonumber\\ 
&\hspace{.5cm}e^{i \d\x^m P_m} (1+i \tilde \ve^i{}_j L_i{}^j + ...) \,,
\end{align}
where we used $\x'=\x+\delta \x$. Since $\d\x^m$ and $\tilde
\ve^i{}_j$ are infinitesimal, the r.h.s.\ of (\ref{sghs}) becomes
\begin{align}
{\rm r.h.s.}=1+ i \d\x^i P_i + i \tilde \ve^i{}_j L_i{}^j + ...\, .  
\end{align}
We are interested in all commutators of the l.h.s.\ of
(\ref{sghs}) which close on $P_i$. They can be found by applying the
Baker-Hausdorff formula.  The relevant commutators are
\begin{align}
[-i\x^m P_m, i \ve^i{}_j L_i{}^j]&= i\ve^i{}_j \x^j P_i , 
\nonumber\\
\frac{1}{2!}[-i\x^m P_m, [-i\x^n P_n, i \ve^i{}_{jk} F^1_i{}^{jk}]]
  &=i \ve^i{}_{jk} \x^j  \x^k P_i ,\nonumber\\
& {\rm etc.} ,
\end{align}
since the commutator of $P_i$ with a generator $F^n$ closes on $F^{n-1}$.
The l.h.s.\ is then given by
\begin{align}
{\rm l.h.s.}&=1+ i [\ve^i + \ve^i{}_j \x^j + \ve^i{}_{jk} 
                \x^j \x^k + ...] P_i \nonumber\\ &+i [\ve^i{}_j +\ve^i{}_{jk} 
                 \x^k + ...]  L_i{}^j + ...\,   \nonumber\\
            &\equiv 1+i \ve^i(\xi) P_i +i \ve^i{}_j(\xi)  L_i{}^j + ...\,.
\label{A200}
\end{align}
It is interesting to observe that the breaking of the translations
effectively makes the group parameters of $g$ depend on $\xi^i$,
compare Eq.\ (\ref{A200}) with the definition of $g$ given in 
Eq.~(\ref{geqn}).

Comparing the coefficients of $P_i$, we get
\begin{align} \label{A37}
\delta \x^i = \ve^i + \ve^i{}_j \x^j + \ve^i{}_{jk} 
                \x^j \x^k + ... \equiv \ve^i(\xi)\, ,
\end{align}
i.e.\ the fields $\xi^i$ transform as coordinates.

\subsection{The transformation behavior of~$\omega^i{}_{jk}$} \label{app2}

Let us now consider the coset space $G/H_1=\textit{Diff}\,(n,
\RR)/GL(n,\RR)$. We choose the parameterizations
\begin{align}
  \s = &\,\tilde \s u=  e^{i \x^m P_m } e^{i \omega^m{}_{nr} F^1_m{}^{nr}} u
  , \\ g \approx& 1 + i \ve^i P_i + i \ve^i{}_j L_i{}^j + i
  \ve^i{}_{jk} F^1_i{}^{jk} + ... , \\ h \approx& 1 + i \tilde
  \ve^i{}_j L_i{}^j \, .
\end{align}
Since the element $u$ associated to the generators $F^m$ $(m \geq 2)$
has no influence on the transformation behavior of $\omega^i{}_{jk}$,
we may consider just $\tilde \s$.  In order to obtain the
transformation behavior $\d \omega^i{}_{jk}$, let us again solve 
Eq.~(\ref{A12}). We obtain
\begin{align}
&e^{-i \d\x^m P_m} e^{-i\x^m P_m} (1 + i \ve^i P_i + ...) e^{i\x^m P_m} =
\nonumber\\
&\hspace{.5cm}
e^{i \omega'{}^m{}_{nr} F^1_m{}^{nr}} (1 + i \tilde \ve^i{}_j L_i{}^j)
e^{-i \omega^m{}_{nr} F^1_m{}^{nr}}
\end{align}
or, equivalently, by employing (\ref{A200})
\begin{align}
&e^{-i \omega^m{}_{nr} F^1_m{}^{nr}} (1 + i [\ve^i(\xi)-\d\xi^i] P_i + 
\ve^i{}_j(\xi) L_i{}^j \nonumber\\&\hspace{.5cm}
+ i \ve^i{}_{jk}(\xi) F^1_i{}^{jk} +...) 
e^{i \omega^m{}_{nr} F^1_m{}^{nr}} 
= \nonumber\\&\hspace{.5cm} (1+ i \d \omega^m{}_{nr} F^1_m{}^{nr} ) (1 + i \tilde 
\ve^i{}_j L_i{}^j) \,.\label{A201}
\end{align}
Here we used (Taylor expansion)
\begin{align}
&e^{i \omega'{}^m{}_{nr}F^1_m{}^{nr}} = e^{i \omega{}^m{}_{nr} F^1_m{}^{nr}
+ i \d \omega^m{}_{nr} F^1_m{}^{nr}} \nonumber\\&\hspace{.5cm}
\approx e^{i \omega{}^m{}_{nr} F^1_m{}^{nr}}
+ e^{i \omega{}^m{}_{nr} F^1_m{}^{nr}}i \d \omega^m{}_{nr}F^1_m{}^{nr} \,.
\nonumber
\end{align}
The transformation law (\ref{A37}) follows again. Since both $\d
\omega^m{}_{nr}$ and $\tilde 
\ve^i{}_j$ are infinitesimal, the r.h.s.\ becomes
\begin{align}
{\rm r.h.s.} = 1 + i \tilde \ve^i{}_j L_i{}^j +  i \d  \omega^m{}_{nr} 
F^1_m{}^{nr}\, .
\end{align}
Using the Baker-Hausdorff formula, the l.h.s.\ reads
\begin{align}
&{\rm l.h.s.} = 1 + i \ve^i{}_j(\xi) L_i{}^j + i \ve^i{}_{jk} (\xi)
F^1_i{}^{jk} \nonumber\\&\hspace{.5cm}+ [-i \omega^m{}_{nr} F^1_m{}^{nr}
, i \ve^i{}_j(\xi) L_i{}^j] + ... \, .
\end{align}
By means of the commutator
\begin{align}
[F^1_m{}^{nr}, L_i{}^j] = i \d^j_m F^1_i{}^{nr}
                               - i \d^n_i  F^1_m{}^{jr} 
                               - i \d^r_i  F^1_m{}^{nj} ,
\end{align}
the l.h.s.\ finally becomes 
\begin{align} 
&{\rm l.h.s.} =  
 1 + i \ve^i{}_j(\xi) L_i{}^j + \nonumber\\& i \left[
      \frac{\partial \ve^i}{\partial \x^l} \omega^l{}_{jk}  
    - \frac{\partial \ve^l}{\partial \x^j} \omega^i{}_{lk}
    - \frac{\partial \ve^l}{\partial \x^k} \omega^i{}_{jl}
    + \half \frac{\partial^2 \ve^i}{\partial 
      \x^j \partial\x^k}\right] F^1_i{}^{jk}
    + ... \,.     \nonumber
\end{align}
A comparison of the coefficients of $F^1_i{}^{jk}$ yields the
transformation be\-ha\-vi\-our\mbox{ (\ref{A36})}.

\subsection{The transformation behavior of~$r^i{}_\alpha$} \label{app3}

In this section we choose the coset space $G/H=\textit{Diff}\,(n,
\RR)/SO(1,n-1)$. We choose the elements
\begin{align}
\sigma =& \,\tilde \sigma u=  e^{i \x^m P_m } e^{i h^{mn} T_{mn}} u ,\\
g \approx& 1 + i \ve^i P_i + i \ve^i{}_j L_i{}^j + i
  \ve^i{}_{jk} F^1_i{}^{jk} + ... , \\
h \approx& 1 + i \tilde \ve^{ij} M_{ij}  \, .
\end{align}
We may again ignore the element $u$ associated to generators $F^m (m
\geq 1)$. Now we determine the type of the indices of the coset
parameter $r^i{}_\alpha$.  Eq.\ (\ref{A12}) becomes after a similar
computation which led to (\ref{A201})
\begin{align}
  &(1 + i \ve^i{}_j(\xi) L_i{}^j + ...) e^{i h^{mn} T_{mn}} =
 \nonumber\\&\hspace{.5cm}
 e^{i (h^{mn}+\d h^{mn}) T_{mn}}  (1 + i \tilde \ve^{ij}
  M_{ij} ) \,. 
\end{align}

In the following we will make use of the Eqs.\ (A.9)-(A.11) \mbox{in
\cite{Lope}}:
\begin{align} 
  e^{i (h^{mn} +\d h^{mn}) T_{mn}}=e^{i h^{mn}T_{mn}} \left[
    1+i(r^{-1})_i{}^\alpha \d r^{i \beta} T_{\alpha\beta}  \right]
  \label{A42}\\ e^{-i h^{mn}T_{mn}} \sigma^{ij} T_{ij} e^{i h^{mn}T_{mn}}
  =(r^{-1})_i{}^\alpha \sigma^i{}_j r^{j\beta} (T_{\alpha\beta}+ M_{\alpha\beta})
  \label{A43}\\ e^{-i h^{mn}T_{mn}} \tau^{ij} M_{ij} e^{i
    h^{mn}T_{mn}} =(r^{-1})_i{}^\alpha \tau^i{}_j r^{j\beta} (T_{\alpha\beta}+
  M_{\alpha\beta})\label{A44},
\end{align}
where $\sigma^{ij}$ and $\tau^{ij}$ are arbitrary tensors.
We lift and lower indices with the Minkowski metric $\eta_{\alpha\beta}$.   

Using (\ref{A43}) and (\ref{A44}) the l.h.s.\ becomes
\begin{align}
{\rm l.h.s.}= 1 + i (r^{-1})_i{}^\alpha \ve^i{}_j(\xi)  r^{j \beta} 
( M_{\alpha\beta}+T_{\alpha\beta}) + ...,
\end{align}
while with help of (\ref{A42}) the r.h.s.\ reads
\begin{align}
{\rm r.h.s}= 1+ i (r^{-1})_i{}^\alpha \d r^{i \beta} T_{\alpha\beta}
 + i \tilde \ve^{\alpha\beta} (M_{\alpha\beta} + T_{\alpha\beta}) +...\, .
\end{align}
The comparison of the coefficients of $T_{\alpha\beta}$ shows that $r^{i \alpha}$
transforms as
\begin{align}         \label{A202}
\d r^{i \alpha} = \frac{\partial \ve^i}{\partial \xi^j} r^{j \alpha} - \tilde \ve_\beta{}^\alpha 
r^{i \beta} \,.
\end{align}
This well-known result is also obtained in \cite{Bori, Isha}.  It says
that the first (latin) index of $r^i{}_\alpha$ transforms covariantly
while the second (greek) one is a Lorentz index. This justifies a
posteriori the use of different types of indices for $r^i{}_\alpha$.

%%%%%%%%%%%%%%%%%%%%%%%%%%%%%%%%%%%%%%%%%%%%%%%%%%%%%%%%%%%%%%%%%%%%%
\section{Connection 1-form for $\mathbf{G/H={Diff}\,(n,\RR)/SO(1,n-1)}$}
\label{sec1form}
%%%%%%%%%%%%%%%%%%%%%%%%%%%%%%%%%%%%%%%%%%%%%%%%%%%%%%%%%%%%%%%%%%%%%

In the following we calculate the translational part $\vartheta^i$ and
the $GL(n, \mathbb{R})$ part $\Gamma_i{}^j$ of the connection $\Gamma$
in case when $H_2=SO(1, n-1)$. Then an element of $G/H$ reads
$\sigma\equiv e^{i \xi^i P_i} e^{i h^{ij} T_{ij}} e^{i \omega^i{}_{jk}
F^1_i{}^{jk} } u$, where $u$ is an element of the group spanned by
$F^n (n \geq 2)$.

Let us first calculate the simpler case when $G/H$ is just spanned by
$P_i$ and $T_{ij}$ with the elements $\tilde \sigma=e^{i \xi^i P_i} e^{i
  h^{ij} T_{ij}}$. Then the nonlinear connection $\Gamma$ becomes
\begin{align}
 &\Gamma = \tilde \sigma^{-1} d \tilde \sigma \\
          &=e^{-i h^{ij} T_{ij} } e^{-i\xi^i P_i}
            [ (d e^{i \xi^i P_i}) e^{i h^{ij} T_{ij}}
            + e^{i \xi^i P_i} d e^{i h^{ij} T_{ij}} ]\nonumber\\
          &=e^{-i h^{ij} T_{ij}} (i d\xi^k P_k) 
            e^{i h^{ij} T_{ij}}
          + e^{-i h^{ij} T_{ij}} i dh^{kl} T_{kl}
            e^{i h^{ij} T_{ij}}  \nonumber\\
          &=i d\xi^k P_k + d\xi^k  h^{ij} [T_{ij}, P_k ] 
           -\frac{i}{2} [ h^{ij} T_{ij},[h^{mn}T_{mn},d\xi^k P_k]]
              \nonumber\\
           &\hspace{.5cm}+ (r^{-1})_i{}^\alpha i d h^{ij} r_j{}^\beta 
             (T_{\alpha\beta}+M_{\alpha\beta}) +... \nonumber\\ 
          &=i e^{-h^\alpha{}_i} P_\alpha d\xi^i + i (r^{-1})_i{}^\alpha
            d r^{i \beta} (T_{\alpha\beta}+ M_{\alpha\beta}) \nonumber\\
          &=i (r^{-1})_i{}^\alpha d\xi^i P_\alpha +  \frac{i}{2} 
            \{r^{-1},dr \}^{\alpha\beta} T_{\alpha\beta}  
           + \frac{i}{2} [r^{-1},dr]^{\alpha\beta}
            M_{\alpha\beta} \,. \nonumber
\end{align}
Here we used Eq.~(\ref{A43}) after the fourth equality sign. The
projections are thus given by
\begin{align}
\tilde \vartheta^\alpha &\equiv  (r^{-1})_i{}^\alpha d\xi^i \,,\\
\tilde \Gamma^{(\alpha\beta)} &\equiv \frac{1}{2} \{r^{-1},dr \}^{\alpha\beta}\,,\\
\tilde \Gamma^{[\alpha\beta]} &\equiv \frac{1}{2} [r^{-1},dr]^{\alpha\beta} .
\end{align} 
Now, consider $\sigma=\tilde \sigma \tilde u$ with $\tilde u\equiv e^{i \omega^i{}_{jk} 
F^1_i{}^{jk}} u$. The connection becomes
\begin{align}
\sigma^{-1}d\sigma&= (\tilde \sigma \tilde u)^{-1} d (\tilde \sigma \tilde u) =
 \tilde u^{-1} 
\tilde \sigma^{-1} [d\tilde \sigma \tilde u + \tilde \sigma d \tilde u] \\
&= \tilde u^{-1} 
[ \tilde \sigma^{-1} d \tilde \sigma] \tilde u 
+ \tilde u^{-1} d \tilde u \nonumber\\
   &= e^{-i \omega^i{}_{jk} F^1_i{}^{jk} } [ i \tilde \vartheta^\delta 
      P_\delta 
    + i \tilde \Gamma^{(\alpha\beta)} T_{\alpha\beta}  \nonumber\\
   &\hspace{0.4cm}+ i \tilde \Gamma^{[\alpha\beta]} M_{\alpha\beta} ]
      e^{i \omega^i{}_{jk} F^1_i{}^{jk} } + O(F^1) \nonumber\\
   &= i \tilde \vartheta^\alpha P_\alpha 
    + i \tilde \Gamma^{(\alpha\beta)} T_{\alpha\beta} 
    + i \tilde \Gamma^{[\alpha\beta]} M_{\alpha\beta} \nonumber\\
   &\hspace{0.4cm}
     + [F^1_i{}^{jk}, P_\delta] \tilde \vartheta^\delta \omega^i{}_{jk} 
    + O(F^1) \nonumber\\
   &= i \tilde \vartheta^\alpha P_\alpha 
    + i \tilde \Gamma^{(\alpha\beta)} T_{\alpha\beta} 
    + i \tilde \Gamma^{[\alpha\beta]} M_{\alpha\beta} \nonumber\\
  & \hspace{0.4cm}- 2i \omega^\alpha{}_{\beta\gamma} \tilde \vartheta^\gamma L_\alpha{}^\beta + O(F^1) \,. \nonumber
\end{align}
Then the connection 1-forms are
\begin{align}
\vartheta^\alpha &\equiv  (r^{-1})_i{}^\alpha d\xi^i ,\\ \Gamma^{(\alpha\beta)}
&\equiv \frac{1}{2} \{r^{-1},dr \}^{\alpha\beta} - \omega^{\alpha\beta\gamma} \vartheta_\gamma -
\omega^{\beta\alpha\gamma} \vartheta_\gamma , \\ \Gamma^{[\alpha\beta]} &\equiv \frac{1}{2}
[r^{-1},dr]^{\alpha\beta} - \omega^{\alpha\beta\gamma} \vartheta_\gamma + \omega^{\beta\alpha\gamma}
\vartheta_\gamma .
\end{align}

\subsubsection*{An identity}
We finally give an identity which is used in Sec.~\ref{sec23}:
\begin{align} \label{A34}
  &\frac{1}{2} \partial_k g_{ij} r^k{}_\gamma r^i{}_\alpha r^j{}_\beta = \nonumber\\&\hspace{.5cm}
 \frac{1}{2} \partial_k [
  (r^{-1})_i{}^\mu (r^{-1})_j{}^\nu] \eta_{\mu\nu} r^k{}_\gamma r^i{}_\alpha
  r^j{}_\beta \nonumber\\ &= \frac{1}{2} \left[ \partial_k (r^{-1})_i{}^\mu
    \eta_{\mu\beta} r^i{}_\alpha + \partial_k (r^{-1})_j{}^\nu \eta_{\alpha\nu} r^j{}_\beta
  \right] r^k{}_\gamma \nonumber\\ &= - \frac{1}{2} \{r^{-1},
  \partial_k r \}_{\alpha\beta} r^k{}_\gamma = - \nabla_\gamma h_{\alpha\beta}\,.
\end{align}

\section{Physical degrees of freedom of a symmetric connection}

In this appendix we decompose a four-dimensional ($n=4$) symmetric
connection $\Gamma_{ij}{}^k$ with 40 (off-shell) components into
irreducible representations under the Lorentz group and determine the
number of physical (on-shell) degrees of freedom of such a
connection. We then show that the Goldstone metric $g_{ij}$ provides
the exact number of degrees of freedom for a massive
nonmetricity tensor.

A connection $\Gamma_{ij}{}^k$, which is symmetric in its lower indices
$i$ and $j$, can be split into two pieces, each with 20 components:
$\Gamma_{i(jk)}$ and $\Gamma_{i[jk]}$. Under the Lorentz group, the
symmetric part $\Gamma_{i(jk)}$ decomposes into a tracefree and a
trace part,
\begin{align}
\Gamma_{i(jk)}= \hat \Gamma_{i(jk)} +\frac 1{4} \Gamma_i g_{jk} \,\quad
(\Gamma_i \equiv \Gamma_{ik}{}^k) \,,
\end{align}
which correspond to the representations $D^{(\frac{1}{2},
\frac{1}{2})}$ and $D^{(\frac{3}{2}, \frac{3}{2})}$ with 4 and 16
components, respectively. In Young tableau notation this
decomposition can be written as:
\begin{align}
GL(4,\, \RR) \hspace{1.5cm} SO(1,3) \hspace{1.5cm} \nonumber\\
\raisebox{0.45cm}{$\Gamma_{i(jk)}$} \quad 
\begin{picture}(35,20)
      \put(0,10) {\line(1,0){30}}              % 3
      \put(0,20) {\line(1,0){30}}
      \multiput(0,10)(10,0){4} {\line(0,1){10}}  
       \put(11,01) {\bf 20} 
     \end{picture} 
\,\, \supset\,\,\,\,\,\,
\begin{picture}(40,20)
      \put(0,10) {\line(1,0){30}}              % 3
      \put(0,20) {\line(1,0){30}}
      \multiput(0,10)(10,0){4} {\line(0,1){10}}  
      \put(11,01) {\bf 16}   
\end{picture} 
 \oplus\,\,
\begin{picture}(40,20)
      \put(0,10) {\line(1,0){10}}              % 1
      \put(0,20) {\line(1,0){10}}
      \multiput(0,10)(10,0){2} {\line(0,1){10}}  
     \put(3,01) {\bf 4}
     \end{picture} 
\end{align}
The representation $D^{(\frac{3}{2}, \frac{3}{2})}$ describes a spin-3
``particle'' which we refer to as TRITON (prefix ``tri'' for spin
three) in accordance with the corresponding nonmetricity component
TRINOM \cite{MAG}.\footnote{Interacting higher spin theories (spin
$>2$) usually face consistency problems. Since in our case TRITON
decouples, it does not cause problems for the low energy effective
theory (General Relativity). However, consistency of the high energy
theory remains to be shown. We leave this for future research.}

The number of physical degrees of freedom of each of these irreducible
pieces is fixed by the dimension of the same representation
transforming under the little group of the Poincar\'e group
$ISO(1,3)$, which is $SO(2)$ ($SO(3)$) in the case of massless
(massive) representations, respectively.  

Tab.~\ref{tabdof} shows the dimensions of these representations.  The
first line of the table tells us that the metric with 10 (off-shell)
components, which splits into a trace and traceless symmetric part
under $SO(1,3)$, describes a massless (massive) graviton with two
(five\footnote{There is actually a sixth mode with spin~0 coming from
the trace of the metric. In a theory for massive gravity this mode is
not considered to be physical and must be project out by the action.})
physical polarizations. The second line of the table shows the degrees
of freedom of the spin-3 particle TRITON associated with the 16
(off-shell) components of the traceless total symmetric part of the
connection: This particle has two (seven) polarizations in case it is
massless (massive). The last line shows the number of degrees of
freedom for the remaining vector piece of the connection. As usual for
a massless (massive) vector representation, it has two (three) physical
degrees of freedom. 

Similarly, it is possible to show that the antisymmetric part
$\Gamma_{i[jk]}$ has 2 physical polarizations. This gives in total 6
on-shell degrees of freedom for a symmetric connection
$\Gamma_{ij}{}^k$ with 40 off-shell components.

In the Higgs mechanism for the breaking of $GL(4, \RR)$ down to
$SO(1,3)$, the symmetric part of the connection $\Gamma_{i(jk)}$ absorbs
the metric and turns into massive nonmetricity,
\begin{align}
Q_{ijk} \equiv 2\Gamma'_{i(jk)}=\partial_i g_{jk} -2\Gamma_{i(jk)} \,, 
\end{align}
cf.\ Eq.~(\ref{redef}). Here the 5 d.o.f.\ of the Goldstone graviton
are ``eaten'' by the spin-3 particle TRITON. (The sixth mode of the
graviton is absorbed by the spin-1 particle associated with the trace
$\Gamma_i$ in (C1).) TRITON becomes massive and decouples at low
energies.

Let us compare the number of on-shell degrees of freedom before and
after this process. Before the condensation the metric is
tachyonic\footnote{Recall that the metric $g_{ij}$ is defined in terms
of the traceless part of the Higgs field $\varphi_{ij}$,
cf.~Eq.~(\ref{met}), which has a negative mass squared.} and has
$5+1=6$ d.o.f., while $\Gamma_{i(jk)}$ and $\Gamma_{i[jk]}$ are
massless and have $2+2=4$ and $2$ d.o.f. During the breaking the
symmetric part of the connection $\Gamma_{i(jk)}$ absorbs all six
d.o.f.~of the metric and becomes massive with $7+3=10$ d.o.f.  After
integrating out these massive modes, we are left with the $2$
d.o.f.~of the antisymmetric part of the connection. Due to the
inverse Higgs effect these modes are identical to the $2$ d.o.f.~of a massless
graviton. Recall that this part of the connection has become the
metric connection.  As required, the total number of d.o.f.,
$12=6+4+2=2+10$, is preserved.
%\vspace{9cm}\mbox{}

\begin{table}[ht]
\begin{tabular}{cccccc}
field & Y.T. &$SO(1,3)$ & $SO(3)$ & $SO(2)$ & name \\
\hline
$g_{ij}$ &
\begin{picture}(40,15)
      \put(0,0) {\line(1,0){20}}              % 2
      \put(0,10) {\line(1,0){20}}
      \multiput(0,0)(10,0){3} {\line(0,1){10}}  
     \end{picture}  &9(+1) & 5(+1) & 2 & GRAVITON\\
\hline
$\hat \Gamma_{i(jk)}$&
\begin{picture}(40,15)
      \put(0,0) {\line(1,0){30}}              % 3
      \put(0,10) {\line(1,0){30}}
      \multiput(0,0)(10,0){4} {\line(0,1){10}}  
     \end{picture}  &16 & 7 & 2 & TRITON\\
$\Gamma_i$&
\begin{picture}(40,15)
      \put(0,0) {\line(1,0){10}}              % 1
      \put(0,10) {\line(1,0){10}}
      \multiput(0,0)(10,0){2} {\line(0,1){10}}  
     \end{picture}  &4 & 3 & 2 &
\end{tabular}
\caption{On-shell and off-shell degrees of freedom of
the metric and the total symmetric part of the connection.} \label{tabdof}
\end{table}

%\vspace{5cm} \mbox{}

\end{document}